\documentclass[secnumarabic,
               amssymb, 
               nobibnotes, 
               aps, 
               prd, 
               longbibliography,
               amsmath]{revtex4-1}
\usepackage{graphicx}
\usepackage{amsmath}
\usepackage{amsfonts}
\usepackage{amssymb}
\usepackage{bm}
\usepackage{color}
\usepackage{verbatim}
\usepackage{datetime}

\newenvironment{nalign}{
    \begin{equation}
    \begin{aligned}
}{
    \end{aligned}
    \end{equation}
    \ignorespacesafterend
}

\begin{document}
\preprint{}
\title{Sub-grid scale characterization and asymptotic behavior of multi-dimensional upwind schemes for the vorticity transport equations}
\author{Daniel Foti}
\affiliation{Department of Mechanical Engineering, University of Memphis, Memphis, TN, USA}
\author{Karthik Duraisamy}
\affiliation{Department of Aerospace Engineering, University of Michigan, Ann Arbor, MI, USA}
\begin{abstract}
\textbf{
We study the sub-grid scale characteristics of a vorticity-transport-based approach for large-eddy simulations. In particular, we consider a  multi-dimensional upwind scheme for the vorticity transport equations and establish its properties in the under-resolved regime. The asymptotic behavior of key turbulence statistics of velocity gradients, vorticity, and invariants is studied in detail.
Modified equation analysis  indicates that dissipation can be controlled locally via non-linear limiting of the gradient employed for the vorticity reconstruction on the cell face such that low numerical diffusion is obtained in well-resolved regimes and  high numerical diffusion is realized in under-resolved regions. The enstrophy budget highlights the remarkable ability of the truncation terms to mimic the true sub-grid scale dissipation and diffusion. The modified equation also reveals diffusive terms that are similar to several commonly employed sub-grid scale models including tensor-gradient and hyper-viscosity models.  
Investigations on several canonical turbulence flow cases show that large-scale features 
are adequately represented and remain consistent in terms of spectral energy over a range of grid resolutions.  Numerical dissipation in under-resolved simulations is consistent and can be characterized by diffusion terms discovered in the modified equation analysis.
A minimum state of scale separation 
 necessary to obtain asymptotic behavior 
 is characterized using metrics such as effective Reynolds number and effective grid spacing.  Temporally-evolving jet simulations, characterized by large-scale vortical structures, demonstrate that high Reynolds number vortex-dominated flows are captured when criteria is met and necessitate diffusive non-linear limiting of vorticity reconstruction be employed to realize accuracy in under-resolved simulations.
 }   
\end{abstract}
\maketitle
%
%
%
\section{Introduction}
\indent Coherent structures are prominent in a wide range of turbulent flows including jets \cite{yule1978large, akselvoll1996large}, wakes \cite{foti2016wake, foti2018wake}, and atmospheric flows \cite{guala2011interactions, fang2015large}. The critical characteristics of such flows are defined to a large extent by the dynamics of dominant coherent structures, pointing to  large-eddy simulation (LES) as an ideal candidate model. In classical LES, a scale-separation is performed typically via a low-pass filter, and a sub-grid scale model is imposed to represent the impact of the unresolved scales on the resolved scales. In practice, the sub-grid scale model provides a pathway for energy to dissipate from the resolved scales because physical dissipation mechanisms (i.e., viscous dissipation at the Kolmogorov scale) are unresolved. However, the accuracy of LES at high Reynolds numbers is significantly influenced by discretization errors~\cite{kravchenko1997effect,chow2003further,freitag2006improved,toosi2020towards}, as these errors can be of a similar magnitude as the sub-grid scale model in practical scenarios.
Despite this fundamental obstacle, functional~\cite{smagorinsky1963general,germano1991dynamic,bardina1980improved,lilly1992proposed,clark1979evaluation,bou2005scale, yu2017scale, chapelier2018coherent} and structural~\cite{schneider2005coherent, ouvrard2010classical, parish2017dynamic,parish2017non,geurts1997inverse,zhou2018structural,kraichnan1970diffusion} LES sub-grid scale models are actively developed and applied to practical flows with varying degrees of success. \\
\indent An alternate view eschews the explicit modeling of sub-grid scales and instead focuses on tailoring numerical dissipation in the underlying discretization errors. Monotone integrated LES (MILES), first proposed by Boris \emph{et al.}~\cite{boris1992new}, utilizes functional reconstruction of the convective fluxes in a monotonicity-preserving finite volume scheme to integrate the effects of the sub-grid scale on the resolved scale.  Results from similar numerical methods also were observed by Youngs~\cite{youngs1991three} around the same time. A more broadly-defined methodology, referred to as implicit LES (ILES), requires a careful consideration of  discretization errors. Typically, ILES employs a certain class of numerical methods, most notably monotonic upwind finite volume methods such as piece-wise parabolic~\cite{gathmann1993numerical}, MPDATA~\cite{smolarkiewicz1998mpdata}, total variation diminishing~\cite{thornber2007implicit} and flux-corrected transport~\cite{wachtor2013implicit}, which have implicit dissipation mechanisms~\cite{boris1992new, margolin2002rationale, margolin2005design}. In practice, ILES methods are employed on inertially dominated (e.g., high Reynolds number) dynamics and regularize the under-resolved scales  similar to the way in which shock-capturing, non-oscillating finite volume schemes use weak solutions and satisfy the entropy condition.

Thorough analysis of the schemes reveals the errors are often similar in form to certain sub-grid scale models~\cite{margolin2006modeling}.   
However, the form of the discretization errors does not necessarily has to be similar to a sub-grid scale model provided the dissipation acts on the high wavenumber range.  Drikakis \emph{et al}~\cite{drikakis2009large}  characterized the effect of numerical resolution on the dissipation into two categories: a) When the solution is well-resolved and an adequate distinction between the start of the inertial range and the dissipation scales exists, the numerical dissipation should not influence the large scales, and the eddies interacting with the largest scales should be reasonably resolved.  The separation of scales should not depend on the form of the dissipation; b) The second category involves inadequate numerical resolution, which is often confronted in engineering applications.  Given an inadequate separation between the large and dissipative scales,  the numerical scheme should be designed to mimic the impact of dissipation on the large scales.    
Many ILES solution fall under this category including isotropic turbulence~\cite{thornber2007implicit, zhou2014estimating, hickel2006adaptive}, geophysical flows~\cite{smolarkiewicz1998mpdata, margolin1999large}, jets~\cite{fureby2002large, boris1992new, grinstein2007flux} and channel flows~\cite{grinstein2007flux}. \\
\indent  In many vortex-dominated problems, especially those in which the vorticity distribution is compact~\cite{hussain1986coherent,papamoschou1993vortex,duraisamy2008evolution} the vorticity-velocity formulation of the Navier--Stokes equations has the potential to be advantageous in comparison to the pressure-velocity form~\cite{koumoutsakos1995high,foti2018multi}. Despite the use of a wide array of Eulerian~\cite{brown2005efficient, whitehouse2014innovative}, Lagrangian~\cite{koumoutsakos2005multiscale, winckelmans1996priori} and mixed numerical implementations~\cite{cottet1990particle, ould2001blending} of the vorticity transport equations,  modeling of unresolved dynamics remains an outstanding issue.
Further, while ILES has been prominently utilized in the context of the velocity-pressure formulation, the authors are not aware of literature investigating implicit sub-grid scale characterizations for schemes using the vorticity-velocity formulation.  The goal of this work is to explore how tailored numerical schemes of vorticity-velocity formulation are suitable and can be characterized for ILES.

The vorticity-velocity formulation  of the incompressible vorticity transport equations (VTE), obtained by applying the curl operator to the momentum equations.  While there is an increase to six (three vorticity and three velocity) variables compared to four (three velocity and one pressure) variables in three dimensions, the formulation is advantageous in flows with compact vorticity distributions typical of  vortex-dominated flow, as the potential flow regions do not need to be part of the computational domain. In Eulerian methods,  accurate boundary conditions can be developed~\cite{rennich1997numerical,duraisamy2008evolution,koumoutsakos1995high}, allowing for efficient simulations in a compact domain. Further, the Poisson equation for pressure, normally a stiff equation, is replaced by the kinematic velocity-vorticity relationship.

Numerical schemes \cite{zhao2011hybrid, stone2010rotor, koumoutsakos2005multiscale} using the Lagrangian description of the flow field have seen considerable success and have applied the LES methodology through development of vorticity sub-grid scale models \cite{winckelmans1996priori, mansfield1998dynamic}.  
Both Refs. \cite{winckelmans1996priori, mansfield1998dynamic} introduced eddy-viscosity type sub-grid scale models for the vorticity-velocity formulation using the Lagrangian description.  
However, Lagrangian frameworks are not optimal for a wide range of turbulent flows. Furthermore, \emph{a priori} tests of forced homogeneous, isotropic turbulence in Ref. \cite{mansfield1998dynamic} reveal that both vorticity convection and vortex stretching contribute to sub-grid scale dissipation, but the dissipation due to vortex stretching is inadequately captured by the sub-grid scale model.  On the other hand, relatively few Eulerian vorticity numerical schemes have been developed \cite{brown2005efficient, meitz2000compact, davies2001novel}.  Recently developed upwind finite volume methods for the VTE \cite{whitehouse2014innovative, parish2016generalized, foti2018multi} have been shown to efficiently capture and preserve vortical structures with relatively coarse resolution (a few grid cells spanning  vortex cores). Consistent integration of both the vorticity convection and vortex stretching terms is a prominent feature of schemes in Refs. \cite{parish2016generalized,foti2018multi}.  While these schemes fit into the philosophy of ILES, an understanding of their sub-grid scale behavior has not been established in the context of turbulent flows. \\ 
\indent 
Margolin \emph{et al}. \cite{margolin2006modeling} used  modified equation analysis to examine leading order diffusive and dispersive error terms.  Based on a Taylor series representation, the modified equation is the effective partial differential equation satisfied by the numerical method.
Their work specifically addresses finite volume approaches, and thus volume-averaged filtering. 
The leading terms in the modified equation are used to elucidate the dissipative effects on the large resolved scales.  As the numerical resolution of the scheme increases, the dissipation should focus on a narrowing range of high wavenumbers and not impact the large scales.  
Several studies \cite{zhou2014estimating, thornber2007implicit} emphasized the calculation of an effective viscosity or Reynolds number for a given grid resolution with resolved features.  Since the filter and dissipation are based on the grid resolution, it is imperative to understand whether there is a sufficient separation of resolved and dissipation scales. A minimum state of scale separation is necessary to reproduce high Reynolds numbers flows with asymptotic turbulence statistics~\cite{zhou2014estimating,chamorro2012reynolds}. 
In this work, we will  examine the utility of a finite volume scheme for the VTE \cite{parish2016generalized, foti2018multi}, characterizes its  ILES properties  and provide a methodology to estimate effective Reynolds numbers.  
This particular formulation has a number of features pertinent for ILES:  Consistent integration of  vorticity convection and vortex stretching terms, promoting stability and accuracy and the ability to represent and preserve sharp gradients. 
Rigorous characterization of the sub-grid scale behavior in canonical turbulent flows enables the use of VTE-based schemes in complex vortex-dominated turbulent flows such as in rotorcraft and wind turbine wakes.\\
\indent We begin the study by introducing the governing and filtered equations in Sec. \ref{sec:govern}.  We provide a description of the numerical scheme and analyze the implicit sub-grid scale model in Secs. \ref{sec:scheme} and \ref{sec:mea}, respectively. We detail the results of numerical experimentation of canonical turbulence flows in Sec. \ref{sec:results}.  Finally, we conclude our work and provide details for future studies in Sec. \ref{sec:conclusions}. 
%
%
%
%
\section{Governing and Filtered Equations}\label{sec:govern}
\indent  We employ the vorticity-velocity formulation of the Navier-Stokes equations obtained by applying the curl operator to the incompressible mass and momentum equations. In compact index notation, the unsteady, three-dimensional incompressible VTE are as follows ($i,j=1,2,3$):
\begin{equation}
   \frac{\partial \omega_i}{\partial t} +
   \frac{\partial} {\partial x_j} \left ( u_j \omega_i - u_i \omega_j  \right ) = 
   f_i  + \frac{1}{Re} \frac{\partial^2 \omega_i}{\partial x_j \partial x_j},
   \label{eqn:vte}
\end{equation} 
where $u_i$ is the velocity and $\omega_i = \epsilon_{ijk} \partial u_k / \partial x_j$ is the vorticity, the curl of the velocity where $\epsilon_{ijk}$ is the Levi-Civita tensor.  $f_i$ is the curl of the body force and $Re = \frac{U L}{\nu}$ is the Reynolds number defined by a characteristic velocity scale $U$, characteristic length $L$, and kinematic viscosity $\nu$. 
The inviscid fluxes in Eqn. (\ref{eqn:vte}) can be rewritten in quasi-linear form as follows:
\begin{equation}
   \frac{\partial \omega_i} {\partial t} + A_1 \frac{\partial}{\partial x_1}  \omega_i + A_2 \frac{\partial}{\partial x_2} \omega_i + A_3 \frac{\partial}{\partial x_3} \omega_i= 0,
   \label{eqn:vte_A}
\end{equation}
where the eigenvalues of matrix $A_1$ are $\lambda_1 = \{ 0, u_1, u_1 \}$, $A_2$ are $\lambda_2 = \{ u_2, 0, u_2 \}$, $A_3$ are $\lambda_3 = \{ u_3, u_3, 0 \}$.  The eigenvector matrices $R_i$ for $A_i$ are 
\begin{equation}
R_{1}  = 
\begin{bmatrix}
   u_1 & 0 & 0 \\ 
   u_2 & 1 & 0 \\ 
   u_3 & 0 & 1 \\
\end{bmatrix},
\: 
R_{2}  = 
\begin{bmatrix}
   1 & u_1 & 0 \\ 
   0 & u_2 & 0 \\ 
   0 & u_3 & 1 \\
\end{bmatrix},
\: 
R_{3}  = 
\begin{bmatrix}
   1 & 0 & u_1 \\ 
   0 & 1 & u_2 \\ 
   0 & 0 & u_3 \\
\end{bmatrix}.
\end{equation}
These equations are degenerate in a hyperbolic sense. As an example, if any component of $u$ equals zero, then the eigenvectors are linearly dependent. This presents difficulties for construction of stable upwind schemes.  This is similar to ideal magneto-hydrodynamics, for which Godunov~\cite{godunov1972symmetric} suggested the addition of a symmetrizing term. 
Following this idea and Ref.~\cite{parish2016generalized}, an additional term is included to stabilize the governing equations. The equations are thus modified in the form
\begin{equation}
   \frac{\partial \omega_i}{\partial t} +
   \frac{\partial} {\partial x_j} \left ( u_j \omega_i - u_i \omega_j  \right ) + u_i \frac{\partial \omega_j} {\partial x_j} = 
   f_i  + \frac{1}{Re} \frac{\partial^2 \omega_i}{\partial x_j \partial x_j},
   \label{eqn:vte_conserv_grp}
\end{equation} 
Included with the VTE due to the incompressible assumption, is the solenoidality of both the velocity and vorticity fields: 
\begin{equation} 
   \frac{\partial u_i}{\partial x_i} = \frac{\partial \omega_i}{\partial x_i} = 0.
   \label{eqn:div_free}
\end{equation}
\indent We refer to Eqn. (\ref{eqn:vte_conserv_grp}) as the modified VTE because the final term on the LHS is an additional term, which is proportional to the divergence of vorticity.  Analytically, the term is zero  (Eqn. (\ref{eqn:div_free})), but numerically, this  stabilizes the hyperbolic system of equations.  The modification in Eqn. (\ref{eqn:vte_conserv_grp}) stabilizes the equations with eigenvalues $\lambda_i = \{ u_i, u_i, u_i \}$ and the eigenvectors of the canonical basis vector for $\mathbb{R}^3$ \cite{parish2016generalized}.  
A similar modification is used in the magneto-hydrodynamic equations to enforce a divergence-free magnetic field \cite{powell1994approximate}, however, in our context we employ the modification for numerical stability. \\ 
\indent The vorticity-velocity formulation in three dimensions has six unknowns.  While the three vorticity variables are obtained through the VTE, a supplemental equation is needed to obtain the velocity induced by the vorticity.  The vorticity-velocity relationship is written in the form of a Poisson equation as follows:
\begin{equation}
   \frac{\partial^2 u_i}{\partial x_j \partial x_j} = - \epsilon_{ijk} \frac{\partial \omega_k}{\partial x_j}.
   \label{eqn:poisson}
\end{equation}
%
%
%
%
%
%
\subsection{The Filtered vorticity transport equations}\label{sec:filter}
\indent By spatial filtering the VTE in Eqn. (\ref{eqn:vte_conserv_grp}), we obtain the following filtered VTE: 
\begin{equation}
   \frac{\partial \widetilde{\omega}_i}{\partial t} +
   \frac{\partial} {\partial x_j} \left ( \widetilde{u}_j \widetilde{\omega}_i - \widetilde{\omega}_j \widetilde{u}_i \right ) + \widetilde{u}_i \frac{\partial \widetilde{\omega}_j} {\partial x_j} = \widetilde{f}_i + 
   \frac{1}{Re} \frac{\partial^2 \widetilde{\omega}_i}{\partial x_j \partial x_j} - \frac{\partial \tau_{ij}}{\partial x_j},
   \label{eqn:vte_les}
\end{equation}
where $\widetilde{\cdot}$ indicates the spatial filtering or a resolved quantity and the $\tau_{ij}$ is the sub-grid scale (SGS) vorticity stress due to the filtering operation.  We note that filtering is not performed through an explicit filter function, and we assume that the filtering operation and the derivatives commute.  The grid cell of the mesh at a size $\Delta$ is an implicit physical-space sharp cut-off filter where the velocity and vorticity fluctuations can only be resolved at a size greater than $\Delta$.  The SGS torque, the divergence of the SGS vorticity stress, which accounts for the unresolved velocity and vorticity fluctuations, is defined as 
\begin{equation}
   \frac{\partial \tau_{ij}}{\partial x_j} = 
   \frac{\partial}{\partial x_j} \left [ \left ( \widetilde{u_j \omega_i} - \widetilde{u}_j \widetilde{\omega}_i \right ) -
                                         \left ( \widetilde{u_i \omega_j} - \widetilde{u}_i \widetilde{\omega}_j \right ) \right ]
   + \left [ \widetilde{u_i \frac{\partial \omega_j} {\partial x_j}} - \widetilde{u}_i \frac{\partial \widetilde{\omega}_j} {\partial x_j}\right ].
   \label{eqn:sgs_tensor}
\end{equation}
The SGS torque term is composed of three different terms: 1.) the vorticity transport by unresolved velocity fluctuations, 2.) the unresolved vortex stretching, and 3.) the unresolved contributions for the vorticity divergence modification in Eqn. (\ref{eqn:vte_conserv_grp}).  Unlike previous LES schemes for the VTE \cite{winckelmans1996priori, mansfield1998dynamic} where there is not an additional vorticity divergence term, the SGS vorticity stress tensor is not purely anti-symmetric.  Here, we attempt to rearrange the SGS terms.
The first two terms in Eqn. (\ref{eqn:sgs_tensor}) can be combined into a single anti-symmetric tensor as follows:   
\begin{equation}
   \left (\tau_{ij} \right )_a =  \left ( \widetilde{u_j \omega_i} - \widetilde{u}_j \widetilde{\omega}_i \right ) -
                                         \left ( \widetilde{u_i \omega_j} - \widetilde{u}_i \widetilde{\omega}_j \right ).
   \label{eqn:sgs_anti}
\end{equation}
However, the third term introduced through the vorticity divergence modification cannot be readily decomposed into an SGS vorticity stress tensor and as such it remains a separate term of the SGS torque:
\begin{equation}
   \left ( \frac{\partial \tau_{ij}}{\partial x_j} \right )_d = \widetilde{u_i \frac{\partial \omega_j} {\partial x_j}} - \widetilde{u}_i \frac{\partial \widetilde{\omega}_j} {\partial x_j}.
   \label{eqn:sgs_sym}
\end{equation}
However, through manipulation, this term can be rewritten as follows:
\begin{equation}
   \frac{\partial \left ( \tau_{ij} \right )_m}{\partial x_j} = \left ( \frac{\partial \widetilde{\omega_j u_i}}{\partial x_j} - \frac{\partial \widetilde{\omega_j}\widetilde{u_i}}{\partial x_j} \right ) - \left ( \widetilde{\omega_j s_{ij}} - \widetilde{\omega_j} \widetilde{s_{ij}} \right ),
\end{equation}
where $s_{ij} = \frac{1}{2} ( \partial_j u_i + \partial_i u_j )$ is the strain rate.  
We can obtain another two portions of  the SGS terms.  The first divergence modification  contains terms that are similar to the third and fourth terms in Eqn. (\ref{eqn:sgs_anti}):
\begin{equation}
   \left (\tau_{ij} \right )_d = \widetilde{u_i \omega_j} - \widetilde{u}_i \widetilde{\omega}_j
   \label{eqn:sgs_d}
\end{equation}
While the other term contains the strain rate:
\begin{equation}
   \frac{\partial \left ( \tau_{ij} \right )_s}{\partial x_j} = - \left ( \widetilde{\omega_j s_{ij}} - \widetilde{\omega_j} \widetilde{s_{ij}} \right ).
   \label{eqn:sgs_s}
\end{equation}
The practical purpose of these terms in the LES methodology is to provide a pathway of energy transfer between the resolved  scales to the unresolved (sub-grid)  scales, which is enabled by ensuring additional dissipation. 
%
%
\section{Multi-dimensional upwind finite volume}\label{sec:scheme}
\begin{figure}
   \begin{center}
      \includegraphics[width=.8\textwidth]{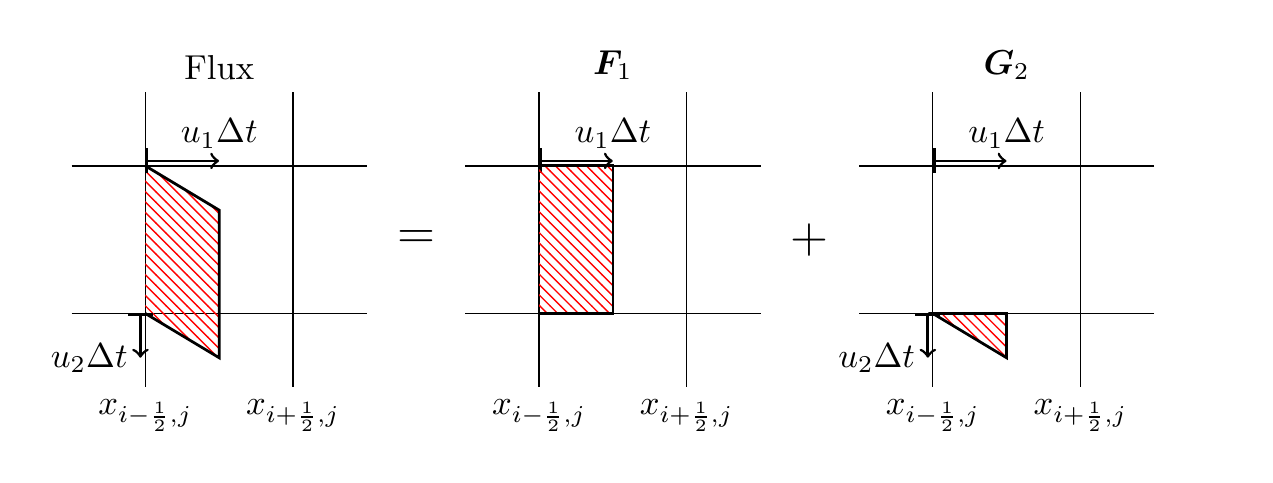}
      \caption{\label{fig:wave} Sketch of multi-dimensional scheme using both normal $\bm{F}_1$ and transverse $\bm{G}_2$ fluxes .}
   \end{center}
\end{figure}

We employ a multi-dimensional upwind finite volume approach to numerically integrate the VTE.  This approach, within the philosophy of the ILES, is similar to Lax-Wendroff-like schemes where upwind differences are corrected by second-order transverse flux corrections to produce a solution that is second-order accurate in space and time. Our multi-dimensional scheme is solved in a series of dimensional sweeps.  The total flux across a cell face in each direction is the sum of several fluxes accounting for both normal and transverse directional fluxes.  A simplified two-dimensional flux is demonstrated in Fig. \ref{fig:wave}.  The updated vorticity is computed as shown (for clarity vectors are shown in bold and $\tilde{\cdot}$ removed from all variables): 
\begin{nalign}
\bm{\omega}^{n+1}_{i,j,k} = \bm{\omega}^n_{i,j,k} &- 
   \frac{\Delta t}{\Delta_{x_1}}  \left [ \left ( (\bm{F}_1^l)^n_{i+\frac{1}{2},j,k} - (\bm{F}_1^r)^n_{i-\frac{1}{2},j,k} \right ) 
                                                - \left ( (\bm{G}_1)^n_{i+\frac{1}{2},j,k} - (\bm{G}_1)^n_{i-\frac{1}{2},j,k} \right ) \right ] \\
                                                &- 
   \frac{\Delta t}{\Delta_{x_2}} \left [ \left ( (\bm{F}_2^l)^n_{i,j+\frac{1}{2},k} - (\bm{F}_2^r)^n_{i,j-\frac{1}{2},k} \right ) 
                                                - \left ( (\bm{G}_2)^n_{i,j+\frac{1}{2},k} - (\bm{G}_2)^n_{i,j-\frac{1}{2},k} \right ) \right ] \\
                                                &- 
   \frac{\Delta t}{\Delta_{x_3}}\left [ \left ( (\bm{F}_3^l)^n_{i,j,k+\frac{1}{2}} - (\bm{F}_3^r)^n_{i,j,k-\frac{1}{2}} \right ) 
                                                - \left ( (\bm{G}_3)^n_{i,j,k+\frac{1}{2}} - (\bm{G}_3)^n_{i,j,k-\frac{1}{2}} \right ) \right ],
   \label{eqn:scheme}
\end{nalign}
where $n$ is the time iteration and $i,j,k$ are the indices of the grid cell centers in three directions.  
The flux functions $\bm{F}^l, \bm{F}^r$ are normal directional (with respect to the cell faces) fluxes and are obtained by solving a generalized Riemann problem developed for the VTE \cite{parish2016generalized}.  In order to account for the fluxes traveling oblique to the cell faces while simultaneously increasing numerical stability and accuracy, transverse fluxes are included. The transverse directional flux functions $\bm{G}$ are computed using the flux-based wave propagation approach \cite{foti2018multi}. All flux functions are stored at the cell faces.  Note that both left and right normal flux functions $\bm{F}_i^l$ and $\bm{F}_i^r$, respectively, are stored at each cell face, while a single transverse flux function $\bm{G}_i$ is stored at each cell face.\\
\indent First, we  detail the normal directional fluxes $\bm{F}_i^l, \bm{F}_i^r$, which are determined by allowing the vorticity to be discontinuous at the cell face.  The sharp discontinuity-capturing scheme can be beneficial for simulating turbulence that is dominated by large coherent structures. In flows such as these, regions that are dominated by coherent structures can be efficiently captured by compact vorticity variables. On the other hand, in under-resolved turbulence regions, the scheme adds additional numerical dissipation (more discussion on this in section \ref{sec:mea}).  \\
\indent The solution to the generalized Riemann problem for vorticity, which is allowed to be discontinuous across the cell face, begins by reconstructing the vorticity stored at the cell center on the cell face.  For simplicity, we will focus on the Riemann problem  at cell face $(i+1/2,j,k)$ and translated to $x_1=0$.  The initial condition for the Riemann problem is
\begin{equation}
   \bm{\omega}_0(x_1, 0) = \left \{
      \begin{array}{@{}ll@{}}
         \bm{\omega}^L+ \left ( x_1 + \frac{1}{2} \Delta_x \right ) \bm{s}^L, & \text{if}\ x_1 < 0\\
         \bm{\omega}^R + \left ( x_1 - \frac{1}{2} \Delta_x \right ) \bm{s}^R, & \text{if}\ x_1 > 0\\
      \end{array} \right., 
   \label{eqn:grp_initial} 
\end{equation}
where $(\cdot)^L$ at ${i,j,k}$,  $(\cdot)^R$ is at ${i+1,j,k}$, $\Delta_x$ is the cell size, $\bm{s}$ is the slope of the vorticity. Achieving second order accuracy via slope reconstruction is beneficial in maximizing the separation of scales for ILES.  In areas of smooth vorticity a second-order accurate central difference can be used, however, as with most second-order accurate discontinuity-capturing schemes, an appropriate slope needs to be used, and a limiter is employed.  A limiter reduces the slope calculation to a first-order accurate difference and effectively adds numerical diffusion.  \\
\indent Integrating the vorticity transport equations, equation (\ref{eqn:vte_conserv_grp}), with the vorticity reconstruction, equation (\ref{eqn:grp_initial}), over the space $ \left [ -\frac{1}{2} \Delta_x, \frac{1}{2} \Delta_x \right ] \times [ 0, \Delta t]$, where $\Delta t$ is the time step, a left moving flux $\bm{F}^l_1$ and right moving flux $\bm{F}^r_1$ are obtained at the cell face as follows ($\bm{F}^{r,l}_2$ and $\bm{F}^{r,l}_3$ follow similarly):
\begin{equation}
   \bm{F}^l_1 = \left \{
      \begin{array}{@{}ll@{}}
         u_1 \hat{\bm{\omega}}^L - \bm{u} \hat{\omega}_1^L + \frac{1}{2} \bm{u} s^L_1 \Delta_x, 
           & \text{if}\ u_1 \ge 0\\
         u_1 \hat{\bm{\omega}}^R - \bm{u} \hat{\omega}_1^R + \frac{1}{2} \bm{u} s^R_1 \Delta_x 
           + \bm{u} \left [ \frac{1}{2} \left ( s^L_1 - s^R_1 \right ) u_1 \Delta t + 
                      \left ( \tilde{\omega}^L_1 - \tilde{\omega}^R_1 \right ) \right ],  
           & \text{if}\ u_1 < 0,
      \end{array} \right.
    \label{eqn:fluxl_grp}
\end{equation}
\begin{equation}
   \bm{F}^r_1 = \left \{
      \begin{array}{@{}ll@{}}
        u_1 \hat{\bm{\omega}}^L - \bm{u} \hat{\omega}_1^L - \frac{1}{2} \bm{u} s^R_1 \Delta_x 
       - \bm{u} \left [ \frac{1}{2} \left ( s^L_1 - s^R_1 \right ) u_1 \Delta t + 
                      \left ( \tilde{\omega}^L_1 - \tilde{\omega}^R_1 \right ) \right ],
            & \text{if}\ u_1 \ge 0\\ 
       u_1 \hat{\bm{\omega}}^R - \bm{u} \hat{\omega}_1^R - \frac{1}{2} \bm{u} s^R_1 \Delta_x,
           & \text{if}\ u_1 < 0,
     \end{array} \right.
   \label{eqn:fluxr_grp}
\end{equation}
where for conciseness, additional vorticity reconstruction variables are created:
\begin{equation*}
    \hat{\bm{\omega}}^L = \bm{\omega}^L + \frac{1}{2} \bm{s}^L \left ( \Delta_x - u_1 \Delta t \right ), \ \
    \hat{\bm{\omega}}^R = \bm{\omega}^R - \frac{1}{2} \bm{s}^R \left ( \Delta_x + u_1 \Delta t \right ),
 \ \
    \tilde{\omega}^L_1 = \omega^L_1 + \frac{1}{2} s^L_1 \Delta_x, \ \
    \tilde{\omega}^R_1 = \omega^R_1 - \frac{1}{2} s^R_1 \Delta_x.
\end{equation*}
\indent Next, we detail the computation of the transverse fluxes. In this scheme we pursue a flux-based wave decomposition introduced in Ref. \cite{bale2003wave}, where the flux difference, $F^l - F^r$, is rewritten as a linear combination of eigenvectors.  This choice is motivated because the solution to the generalized Riemann problem is known, is linearly independent in each direction, and can be used directly without costly manipulations.  In this implementation, the fluxes, $\bm{F}_1^l$ and $\bm{F}_1^r$, only need to be computed once (per time-step), and the transverse fluxes are evaluated with that solution. The flux difference is decomposed into \emph{f}-waves $\mathcal{Z}^p$, the flux wave, as follows:
\begin{equation}
    \bm{F}^l - \bm{F}^r = \sum_{p=0}^m \beta^p r^p \equiv \sum_{p=1}^m \mathcal{Z}^p,
    \label{eqn:fwave}
\end{equation}
and 
\begin{equation}
   \bm{\beta} =  R^{-1} \left ( \bm{F}^l - \bm{F}^r  \right ),
   \label{eqn:flux_strength}
\end{equation}
where the eigenmatrix $R$ is simply the identity matrix \cite{foti2018multi}.  \\
\indent From Eqn. (\ref{eqn:flux_strength}) and the eigensystem, we obtained a simple relationship for the \emph{f}-waves $\mathcal{Z}^p = \bm{F}^l - \bm{F}^r$ exactly.  The relationship allows for the multi-dimensional wave propagation corrections to be implemented equivalently to a three-dimensional advection equation described in Ref. \cite{leveque1996high} if the \emph{f}-waves are normalized by the eigenvalues $\lambda_i$.  \\
\indent In a multi-dimensional problem, the fluxes propagate in the transverse $x_2$- and $x_3$-direction depending on the wave speed given by the eigenvalues of the Jacobian, which are exactly the velocities $u_i$ at the cell face. The transverse fluxes are implemented by using the \emph{f}-waves and transverse velocity at the grid cell $(i,j,k)$ and updating the transverse fluxes in surrounding cells.  The most important transverse fluxes are the two-dimensional fluxes given by the following:
\begin{nalign}
    (\bm{G}_2)^n_{I,J-\frac{1}{2},k} &= - \frac{1}{2} \frac{\Delta t}{\Delta x_1} u_2 \mathcal{Z}_1, \\
    (\bm{G}_3)^n_{I,j,K-\frac{1}{2}} &= - \frac{1}{2} \frac{\Delta t}{\Delta x_1} u_3 \mathcal{Z}_1,
   \label{eqn:transverse_update}
\end{nalign}
where $n$ is the time step iteration and the indices for the grid cell influenced by the transverse flux are given by
\begin{align*} 
   I = \left \{
   \begin{array}{@{}ll@{}} 
      i,   & \text{if} \: u_1>0 \\
      i-1, & \text{if} \: u_1 < 0
   \end{array} \right., \: \:  
   J = \left \{
   \begin{array}{@{}ll@{}} 
      j+1, & \text{if} \: u_2>0 \\
      j,   & \text{if} \: u_2 < 0
   \end{array} \right., \: \:
   K = \left \{
   \begin{array}{@{}ll@{}} 
      k+1, & \text{if} \: u_3>0 \\
      k,   & \text{if} \: u_3 < 0
   \end{array} \right. .
\end{align*}  
\indent Additional transverse fluxes are employed, which account for three-dimensional fluxes as well as higher-order corrections.  For a detailed implementation see Foti and Duraisamy \cite{foti2018multi}. \\
%
%
\section{Modified equation analysis}\label{sec:mea}
\indent In the following, we present the modified equation analysis (MEA) for the multi-dimensional generalized Riemann problem-based upwind finite volume scheme.  MEA analysis was first proposed in Ref. \cite{hirt1968heuristic} and subsequently used to characterize the unresolved scales for ILES~\cite{margolin2002rationale,margolin2006modeling}. The intuition is that in certain classes of numerical methods, the effects of the unresolved scales can be represented by the truncation error of the discretization.  This analysis must be carefully exploited when considering high Reynolds number turbulent flows discretized with upwind finite volume schemes.  Slope limiting with a multi-dimensional scheme yields a complex, nonlinear scheme.  In particular, the dissipation is proportional to  the multi-dimensional interfacial wave jumps. It must also be recognized that in under-resolved flows, the leading order terms may not necessarily be the best approximator of the truncation terms.  \\  
Due to the complexity of including the exact form of the slope limiter employed in the vorticity reconstruction in the scheme, two limiting cases of the modified equation are analyzed: 1.) the vorticity at the cell interface is smooth and a second-order central difference can be employed, and 2.) the vorticity at the cell interface is discontinuous and a limiter switches the slope to a first-order upwind difference.  

\indent The procedure to develop the modified equation begins with expanding the scheme with Taylor series expansions \cite{hirt1968heuristic, margolin2005design}.  For example, the scheme in Eqn. (\ref{eqn:scheme}) contains the following, which can be substituted with a series expansion:
\begin{equation}
   \frac{\omega_i^{n+1} - \omega_i^n}{\Delta t} = \frac{\partial \omega_i}{\partial t} + \frac{\Delta t}{2} \frac{\partial^2 \omega_i}{\partial t^2} + \frac{\Delta t^2}{6} \frac{\partial^3 \omega_i}{\partial t^3} + \mathcal{O}(\Delta t^3),
\end{equation}
where $n$ is the vorticity at $t$ and $n+1$ is the vorticity at $t+\Delta t$.  Similarly, all terms in the flux functions in Eqns. (\ref{eqn:fluxl_grp}), (\ref{eqn:fluxr_grp}), (\ref{eqn:transverse_update}) can be substituted with Taylor series expansions in terms of vorticity as a function of time $t$ or space $\bm{x}$.  The accumulation of all expansions in the scheme can be manipulated to include terms in Eqn. (\ref{eqn:vte_conserv_grp}) as follows:
\begin{equation}
   \frac{\partial \omega_i}{\partial t} +
   \frac{\partial} {\partial x_j} \left ( u_j \omega_i - u_i \omega_j  \right ) + u_i \frac{\partial \omega_j} {\partial x_j} = 
   \nu \frac{\partial^2 \omega_i}{\partial x_j \partial x_j} + T_{ij} ,
\end{equation}
where $T_{ij}$ is the term that includes all second-order accurate and higher terms from the series expansions not included in the VTE.  The remainder is manipulated to substitute temporal derivative terms with spatial terms using the governing equations.  As such, the remainder $T_{ij}$ is a complex function that contains many high order spatial derivatives of the vorticity and velocity.  In the philosophy of ILES, the leading order terms in the expansion term $T_{ij}$ are related to the SGS vorticity stress $R_{ij}$ due to truncation as follow:
\begin{equation} 
    T_{ij} = \frac{\partial}{\partial x_j} R_{ij} + \mathcal{O}(\Delta_x^3). 
\end{equation}
$T_{ij}$ acts as an implicit sub-grid scale model, which is completely dependent on resolved variables.  Note that there is a distinction between the SGS vorticity stress $R_{ij}$ due to truncation and theoretical SGS vorticity stress $\tau_{ij}$ due to filtering.  In what follows we demonstrate that by changing the form of the slope employed in the vorticity reconstruction on the cell face, the modified equation can be tailored. \\ 
\indent First, we examine terms that remain after the VTE is subtracted from the Taylor series expansion of the scheme in smooth regions and a second-order accurate central difference is employed, i.e. $s^l_1 = 1/(2 \Delta_x)\left (\omega(x+\Delta_x) - \omega(x-\Delta_x) \right)$.  In turbulent flows in which we are interested, a smooth spatial region of vorticity often  corresponds to a resolved (or nearly resolved) region  dominated by a large-scale coherent vortical structure.

In the modified equation analysis, the lowest order remaining terms are second-order accurate terms in space and are considered to  have the largest influence on the scheme.  Moreover, these terms can be collected to implicitly provide a model for the SGS torque.   Eqn. (\ref{eqn:mea2}) shows the second-order accurate terms that make up the implicit SGS torque.  For clarity in presentation, only the $x_1$ and $x_2$ directions are shown ($j=1,2$); however, the third $x_3$ direction follows the same form: 
\begin{nalign}
   \frac{\partial R_{1j}}{\partial x_j} &= \Delta_1^2 
       \left (\frac{1}{12}  u_1 \frac{\partial^3 \omega_1}{\partial x_1^3}  -\frac{1}{4}  u_1 \frac{\partial^3 \omega_1}{\partial x_1^3}-
        \frac{1}{8} \frac{\partial^2 u_1}{\partial x_1^2}\frac{\partial \omega_1}{\partial x_1} + 
       \frac{1}{12} u_2 \frac{\partial^3 \omega_1}{\partial x_2^3}  -
        \frac{1}{4}  u_1 \frac{\partial^3 \omega_2}{\partial x_1^3}  - 
        \frac{1}{8} \frac{\partial^2 u_2}{\partial x_2^2}\frac{\partial \omega_1}{\partial x_2} \right ),\\
   \frac{\partial R_{2j}}{\partial x_j} &= \Delta_2^2  
       \left (\frac{1}{12}  u_1 \frac{\partial^3 \omega_2}{\partial x_1^3}  -
        \frac{1}{4}  u_2 \frac{\partial^3 \omega_1}{\partial x_2^3} -
        \frac{1}{8} \frac{\partial^2 u_1}{\partial x_1^2}\frac{\partial \omega_2}{\partial x_1} + 
        \frac{1}{12}  u_2 \frac{\partial^3 \omega_2}{\partial x_2^3} -
        \frac{1}{4}  u_2 \frac{\partial^3 \omega_2}{\partial x_2^3}-
        \frac{1}{8} \frac{\partial^2 u_2}{\partial x_2^2}\frac{\partial \omega_2}{\partial x_2} \right ),
    \label{eqn:mea2}
\end{nalign}
where $\Delta_i$ is the grid cell size in the $i$th direction.
All terms are dispersive (containing odd-ordered derivatives of vorticity) in nature indicating that in smooth regions where a second-order central difference is used, there are no second-order accurate implicit diffusive terms (even-ordered derivatives of vorticity), which can act as numerical routes for implicit SGS dissipation. 
We take a closer look at the diagonal terms ($\partial_1 R_{11}$ and $\partial_2 R_{22}$) of the SGS torque to obtain some insight into the effects of the modification of the vorticity transport equations and the effect of the anti-symmetric and non-symmetric SGS terms (due to Eqn. (\ref{eqn:sgs_anti}) and Eqn. (\ref{eqn:sgs_sym}), respectively).  
   \\
\indent By rewriting terms in Eqn. (\ref{eqn:mea2}) that are associated with the diagonal terms of $R_{ij}$, an anti-symmetric part of the SGS torque associated with Eqn. (\ref{eqn:sgs_anti}), $\partial_j R^a_{ij}$, can be formed and shown to be equal to zero in the following:
\begin{nalign}
  \frac{\partial R^a_{11}}{\partial x_1}  &= \frac{1}{12} \Delta_x^2 u_1 \frac{\partial^3 \omega_1}{\partial x_1^3} -
                                                              \frac{1}{12} \Delta_x^2 u_1 \frac{\partial^3 \omega_1}{\partial x_1^3} = 0,\\
  \frac{\partial R^a_{22}}{\partial x_2} &= \frac{1}{12} \Delta_x^2 u_2 \frac{\partial^3 \omega_2}{\partial x_2^3} -
                                                              \frac{1}{12} \Delta_x^2 u_2 \frac{\partial^3 \omega_2}{\partial x_2^3} = 0.
  \label{eqn:mea2_anti}
\end{nalign}
The remaining leading order terms can collected into the following:
\begin{nalign}
    \frac{\partial R^d_{11}}{\partial x_1}   = -\frac{1}{6} \Delta_2^2 u_1 \frac{\partial^3 \omega_1}{\partial x_1^3} 
                                     -\frac{1}{8} \Delta_1^2\frac{\partial^2 u_1}{\partial x_1^2}\frac{\partial \omega_1}{\partial x_1}, \\
     \frac{\partial R^d_{22}}{\partial x_2}   = -\frac{1}{6} \Delta_1^2 u_2 \frac{\partial^3 \omega_2}{\partial x_2^3} 
                                     -\frac{1}{8} \Delta_2^2\frac{\partial^2 u_2}{\partial x_2^2}\frac{\partial \omega_2}{\partial x_2},
    \label{eqn:mea2_sym}
\end{nalign}
where $\partial_j R^d_{ij}$ is due to the modification term added to the vorticity transport equation. 
Terms associated with the off-diagonals of $R_{ij}$  contain dispersive terms.
This modified equation of the scheme near a smooth vorticity field shows that there is a limiting case that can be used to reduce the amount of numerical dissipation added by the numerical method, which can control the sub-grid scale dissipation and energy transfer.     \\
\indent The next limiting case employs a limiter to reduce the order of the slope in regions in which vorticity gradients may be very large. In numerical simulations of turbulence, large gradients and discontinuities arise in under-resolved regions of the flow where gradients caused by eddies the size of the grid cell or larger cannot be smoothed by smaller scale eddies that are physically present in the flow but are not numerically captured. Towards this end, a forward/backward difference is employed.  For example, the slope of the vorticity is calculated as $s^l_1 = 1/\Delta_x\left (\omega(x_1) - \omega(x_1-\Delta_x) \right)$.
These  eddies transfer energy in a turbulent flow, and the reduction in order of the slope essentially introduces diffusion into the solution in order to maintain non-oscillatory behavior.  The lowest-order terms remaining from the modified equation are shown in the following (again we limit the result to two dimensions for clarity but the $x_3$-direction has the same form):
\begin{nalign}
   \frac{\partial R_{1j}}{\partial x_j} = & 
                 \frac{\Delta_1}{2}  u_1 \frac{\partial^2 \omega_1}{\partial x_1^2} 
               + \Delta_1^2 \left ( -\frac{1}{8}  \frac{\partial^2 u_1}{\partial x_1^2}\frac{\partial \omega_1}{\partial x_1}
               - \frac{1}{6} u_1 \frac{\partial^3 \omega_1}{\partial x_1^3} \right ) +\\
              & \frac{\Delta_1}{2}  u_2 \frac{\partial^2 \omega_1}{\partial x_2^2} 
               + \Delta_1^2 \left (-\frac{1}{8}  \frac{\partial^2 u_2}{\partial x_2^2}\frac{\partial \omega_1}{\partial x_2}
               + \frac{1}{4} \frac{\partial u_2}{\partial x_2}\frac{\partial^2 \omega_1}{\partial x_2^2}
               - \frac{1}{4} \frac{\partial u_1}{\partial x_1}\frac{\partial^2 \omega_2}{\partial x_1^2}
               + \frac{1}{3} u_2 \frac{\partial^3 \omega_1}{\partial x_2^3}
               - \frac{1}{2} u_1 \frac{\partial^3 \omega_2}{\partial x_1^3} \right ),
   \label{eqn:mea1x}
\end{nalign} 
\begin{nalign}
   \frac{\partial R_{2j}}{\partial x_j} = & 
                \frac{\Delta_2}{2}  u_1 \frac{\partial^2 \omega_2}{\partial x_1^2} 
               + \Delta_2^2 \left (-\frac{1}{8}  \frac{\partial^2 u_1}{\partial x_1^2}\frac{\partial \omega_2}{\partial x_1}
               + \frac{1}{4} \frac{\partial u_1}{\partial x_1}\frac{\partial^2 \omega_2}{\partial x_1^2}
               - \frac{1}{4} \frac{\partial u_2}{\partial x_2}\frac{\partial^2 \omega_1}{\partial x_2^2}
               + \frac{1}{3} u_1 \frac{\partial^3 \omega_2}{\partial x_1^3} 
               - \frac{1}{2} u_2 \frac{\partial^3 \omega_1}{\partial x_2^3} \right ) + \\
               & \frac{\Delta_2}{2}  u_2 \frac{\partial^2 \omega_2}{\partial x_2^2} 
               + \Delta_2^2 \left ( -\frac{1}{8}  \frac{\partial^2 u_2}{\partial x_2^2}\frac{\partial \omega_2}{\partial x_2}
               - \frac{1}{6} u_2 \frac{\partial^3 \omega_2}{\partial x_2^3} \right ),
   \label{eqn:mea1y}
\end{nalign} 
which include a single first-order accurate in space term and several second-order accurate terms.   We can readily see that this case includes both diffusive and dispersive terms for vorticity.  This indicates that limiting the slope of the vorticity is necessary to add diffusive terms to include dissipation to solve turbulent flows through an ILES.   
As with the first limiting case, a few of the terms in Eqn. (\ref{eqn:mea1x}) can be rewritten to show a zero diagonal of an anti-symmetric portion of the SGS torque in the following:
\begin{equation}
   \frac{\partial R^a_{11}}{\partial x_1}  = \Delta_1^2 \left ( 
                         \frac{1}{4} \frac{\partial u_1}{\partial x_1}\frac{\partial^2 \omega_1}{\partial x_1^2}
                        -\frac{1}{4}  \frac{\partial u_1}{\partial x_1}\frac{\partial^2 \omega_1}{\partial x_1^2}
                        + \frac{1}{3} u_1 \frac{\partial^3 \omega_1}{\partial x_1^3}
                       -\frac{1}{3}  u_1 \frac{\partial^3 \omega_1}{\partial x_1^3} \right ) = 0.
   \label{eqn:mea1_anti}
\end{equation}
Similarly, terms can be rearranged in $\partial_2 R^a_{22}=0$ and $\partial_3 R^a_{33}=0$.  
The rest of the terms in the diagonals are accumulated in the part of the SGS torque corresponding to the introduction of the vorticity divergence term modification, which includes a first-order accurate diffusive term.    \\
\indent The analysis reveals that each component in the SGS vorticity stress tensor $R_{ij}$ contains diffusive terms.  In the following we select terms from Eqn. (\ref{eqn:mea1x}) to explore how the diffusion is present in the SGS vorticity stress tensor and its similarity to well-known SGS models.  We emphasize that similarities to explicit SGS models are employed to just enhance our understanding of the ILES dissipation.  In complex turbulent flows, the process of limiting the slope and multi-dimensionality can affect the actual leading terms of the modified equation such that they may not precisely mimic the action of explicit SGS models.  We show that three separate diffusive SGS vorticity stress mechanisms are present in the modified equation and can be summed as follows:
\begin{equation}
    R_{ij} = R^g_{ij} + R^h_{ij} + R^t_{ij}.
\end{equation}
The first term is constructed as follows:
\begin{equation}
   \frac{\partial R^g_{12}}{\partial x_2}  = 
                                  -\frac{1}{8} \Delta_2^2 \frac{\partial^2 u_2}{\partial x_2^2}\frac{\partial \omega_1}{\partial x_2}
                                  +\frac{1}{4} \Delta_2^2 \frac{\partial u_2}{\partial x_2}\frac{\partial^2 \omega_1}{\partial x_2^2},
\end{equation}
We can manipulate the terms by introducing another dispersion term and integrating the SGS torque to obtain a single form for the off-diagonal elements in the SGS vorticity stress tensor, which can be written in index notion as follows:
\begin{equation}
     R^g_{ij} = \frac{1}{4} \Delta^2  \frac{\partial u_j}{\partial x_k}\frac{\partial \omega_i}{\partial x_k},
    \label{eqn:mea1_grad}
\end{equation}
where $\Delta$ is the grid spacing in the $i$th direction.
The form of the Eqn. (\ref{eqn:mea1_grad}) can be readily seen as a term similar to tensor-gradient models \cite{clark1979evaluation, borue1998local, vreman1996large} for the sub-grid scale.  These models have been shown to be not overly dissipative and are often paired with an eddy viscosity model \cite{vreman1996large, borue1998local}. They also provide both dissipation and backscatter, an important aspect of turbulence, which is not possible in eddy viscosity-type models.  However, when employed alone, the tensor-gradient model can be unstable \cite{vreman1996large}. \\
\indent An alternate form for the numerical dissipation can be explored with another group of dispersive and diffusive terms as shown in the following:  
\begin{equation}
  \frac{\partial R^h_{12}}{\partial x_2} =  - \frac{1}{2} \Delta_2^2 u_2 \frac{\partial^3 \omega_1}{\partial x_2^3} 
                                    - \frac{1}{4} \Delta_2^2 \frac{\partial u_1}{\partial x_2}\frac{\partial^2 \omega_1}{\partial x_2^2},   
\end{equation}
where these terms again can be manipulated and integrated.  The second numerical dissipation term for the SGS vorticity stress has the form :
\begin{equation}
    R^h_{ij} = -\frac{1}{4} \Delta^2 u_j \frac{\partial^2 \omega_i}{\partial x_k \partial x_k }.
    \label{eqn:mea1_hyper}
\end{equation}
This term has the form that is similar to hyper-viscosity models \cite{winckelmans1996priori, cerutti2000spectral}, which have the form of $(-1)^{n+1} \nabla^{2n} \bm{u}$, where $n>1$.    Ref. \cite{cerutti2000spectral} used the hyper-viscosity model for the SGS stress tensor with $\nabla^2 \widetilde{s}_{ij}$, which was added to a standard Smagorinsky model and motivated by SGS dissipation of enstrophy.  Hyper-viscosity models with increasing $n$ have less effect on dissipation and `bottleneck' effects shown on the turbulence spectra \cite{lamorgese2005direct}.  However, paired with the tensor-gradient model, it may provide less dissipation than is observed with standard mixed models. \\
\indent With these two terms, all of the diffusive terms in the modified equation and SGS torque are accounted except for the single first-order diffusive term, which appears in both the diagonal and non-diagonal elements of the SGS vorticity stress tensor.   This term can be integrated to show another form of the numerical dissipation:  
\begin{equation}
   R^t_{ij} = \frac{1}{2} \Delta u_j \frac{\partial \omega_i}{\partial x_j}.
   \label{eqn:mea1_t}
\end{equation}
These terms show that in the limit of a slope reconstruction with a first-order slope, there are several modes for additional numerical diffusion that can be used to increase the energy transfer from the resolved scales to the sub-grid scales and add dissipation.  \\
\indent The two limiting cases for the present scheme show that the dissipation of the scheme can be controlled using well-designed limiters for specific flows.  The present scheme is developed to conserve coherent vortical structures in flows.  When regions of the flow field are dominated by large coherent structures, which are not subject to the inertial range energy transfer determined by the SGS stresses, the vorticity tends to be smooth and less implicit dissipation is included.  However, in regions of small-scale vortical structures where the flow field is under-resolved and discontinuities in the vorticity are present, the dissipation needs to be increased to account for the energy transfer to the sub-grid scales.   This formulation provides a plausible framework to simulate large-scale features of turbulent flows without providing dissipation at all flow scales, a common problem for Smagorinsky-like explicit methods. 
%
%
%
%
\section{Numerical Experimentation}\label{sec:results}
\indent In this section, we discuss results obtained through numerical experimentation of several canonical flows of turbulence in a periodic box. We investigate a Taylor-Green vortex in Sec. \ref{sec:tg}, forced isotropic turbulence in Sec. \ref{sec:forced}, and a temporally evolving jet in Sec. \ref{sec:jet}.  
All simulations employ the multi-dimensional wave propagation approach described above.
\subsection{Taylor-Green vortex}\label{sec:tg}
\begin{figure}
   \begin{center}
      \includegraphics[width=\textwidth]{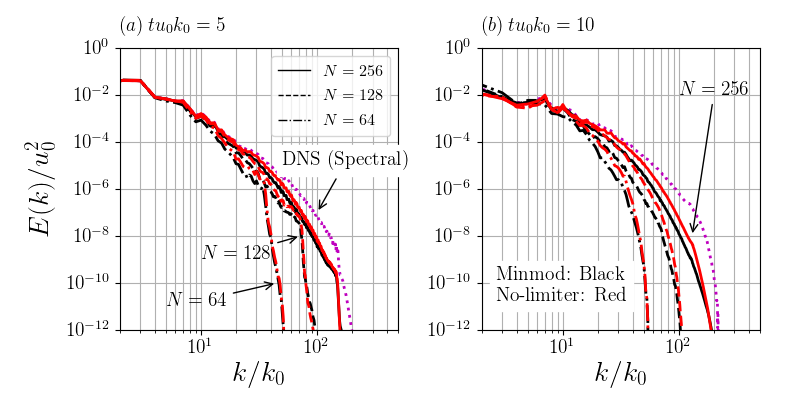}
      \caption{ \label{fig:tg_ee} Energy spectra $E(k)$ of simulations at several under-resolved grid resolutions at $\mathrm{Re}=1600$ at (a) $tu_0k_0=5$ and (b) $tu_0k_0=10$.  Red: no-limiter, Black: minmod limiter. The magenta dotted line is the energy spectra from DNS results.}
   \end{center}
\end{figure}

\begin{figure}
   \begin{center}
      \includegraphics[width=\textwidth]{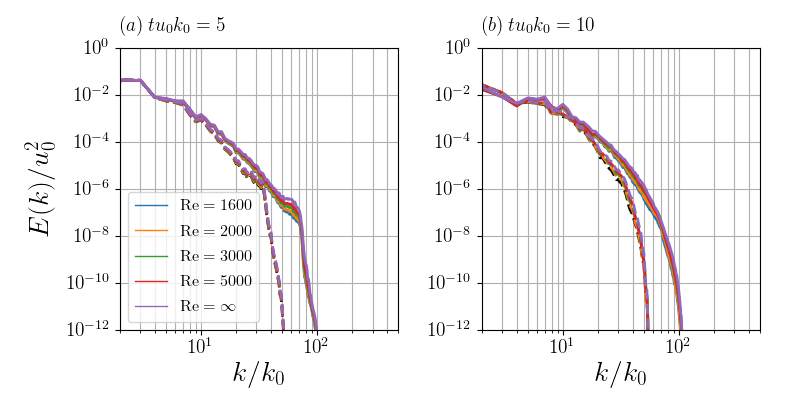}
      \caption{ \label{fig:tg_ee_re} Energy spectra $E(k)$ of simulations at several under-resolved grid resolutions $N=128$ (solid line) and $N=64$ (dashed line) at $\mathrm{Re} = 1600, 2000, 3000, 5000$, and $\infty$ at times (a) $tu_0k_0=5$ and (b) $tu_0k_0=10$. }
   \end{center}
\end{figure}

\begin{figure}
   \begin{center}
      \includegraphics[width=\textwidth]{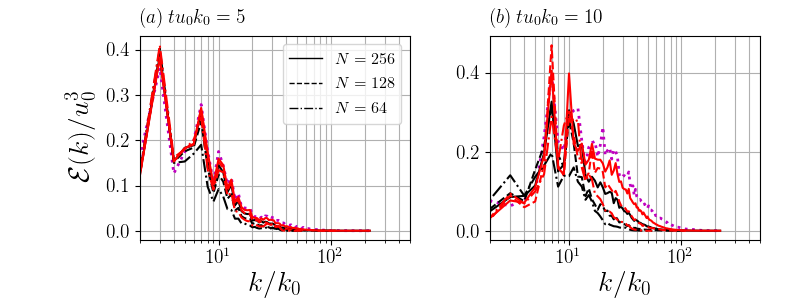}
      \caption{ \label{fig:tg_ens} Dissipation spectra $\mathcal{E}(k)$ from the resolved vorticity of simulations at several under-resolved grid resolutions at $Re=1600$ at (a) $tu_0k_0=5$ and (b) $tu_0k_0=10$. Red: no-limiter, Black: minmod limiter. The magenta dotted line is the dissipation spectra from DNS results.}
   \end{center}
\end{figure}

\begin{figure}
   \begin{center}
      \includegraphics[width=\textwidth]{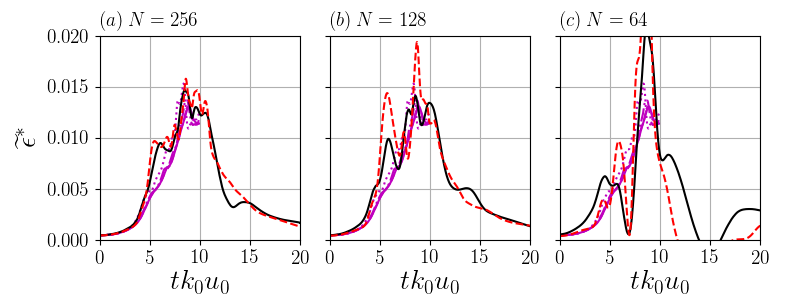}
      \caption{ \label{fig:tg_diss} The filtered dissipation $\widetilde{\epsilon}^*$  as a function of time for (a) $N=256$, (b) $N=128$, and (c) $N=64$. Red dashed line: no-limiter, Black solid line: minmod limiter.  The magenta line is the dissipation from DNS results spatially filtered with a sharp-cutoff spectral filter with several filter widths $\Delta = 2\Delta_x$ (solid line), $4\Delta_x$ (dashed line), and $8\Delta_x$ (dotted line).}
   \end{center}
\end{figure}

\begin{figure}
   \begin{center}
      \includegraphics[width=\textwidth]{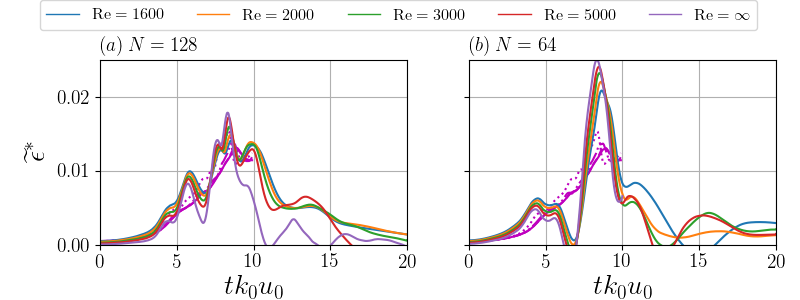}
      \caption{ \label{fig:tg_re_diss} The filtered dissipation $\widetilde{\epsilon}^*$  as a function of time for (a) $N=128$ and (b) $N=64$ at $\mathrm{Re} = 1600, 2000, 3000, 5000$, and $\infty$ . The magenta line is the dissipation from DNS results spatially filtered with a sharp-cutoff spectral filter with several filter widths $\Delta = 2\Delta_x$ (solid line), $4\Delta_x$ (dashed line), and $8\Delta_x$ (dotted line).}
   \end{center}
\end{figure}

\begin{figure}
   \begin{center}
      \includegraphics[width=\textwidth]{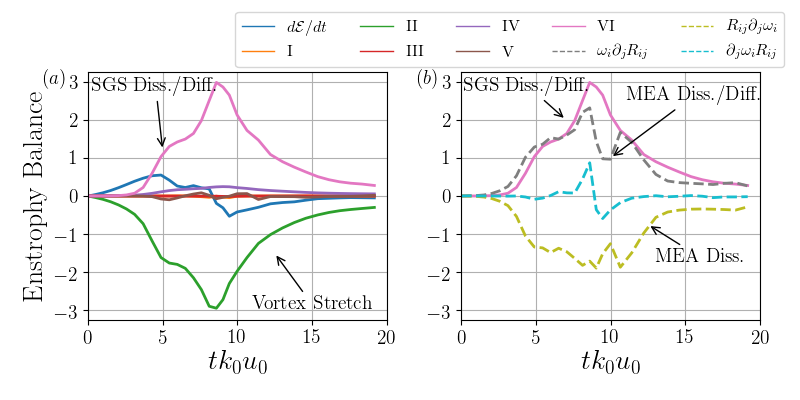}
      \caption{ \label{fig:tg_ens_bal_64} (a) The temporal evolution of terms in the enstrophy transport equation for the $N=64$ test case with minmod limiter. (b) The temporal evolution of the calculated SGS dissipation and diffusion and the temporal evolution of the SGS dissipation and diffusion obtained from the modified equation for the $N=64$ test case with minmod limiter. }
   \end{center}
\end{figure}

\begin{figure}
   \begin{center}
      \includegraphics[width=\textwidth]{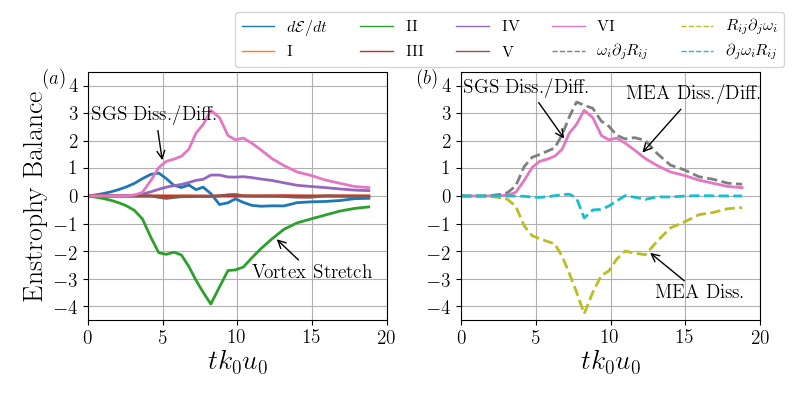}
      \caption{ \label{fig:tg_ens_bal_128} (a) The temporal evolution of terms in the enstrophy transport equation for the $N=128$ test case with minmod limiter. (b) The temporal evolution of the calculated SGS dissipation and diffusion and the temporal evolution of the SGS dissipation and diffusion obtained from the modified equation for the $N=128$ test case with minmod limiter. }
   \end{center}
\end{figure}

\begin{figure}
   \begin{center}
      \includegraphics[width=\textwidth]{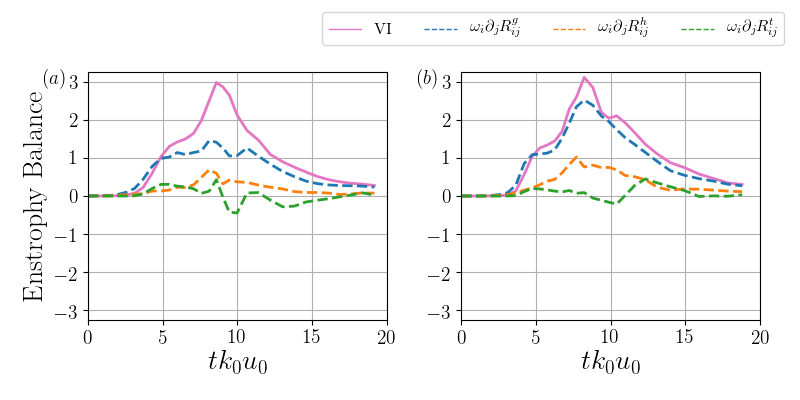}
      \caption{ \label{fig:tg_ind_ens} The temporal evolution enstrophy transport of components of the SGS dissipation and diffusion obtained from modified equation in Eqns. (\ref{eqn:mea1_grad}), (\ref{eqn:mea1_hyper}) and (\ref{eqn:mea1_t}) for (a) $N=64$ and (b) $N=128$ test cases with minmod limiter.}
   \end{center}
\end{figure}

\begin{figure}
   \begin{center}
      \includegraphics[width=\textwidth]{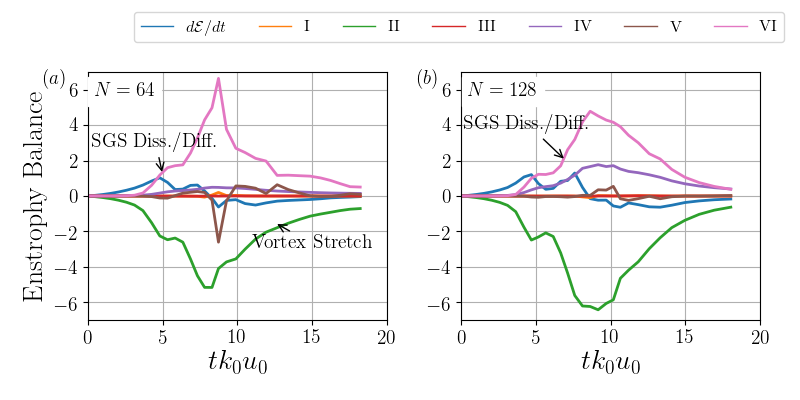}
      \caption{ \label{fig:tg_ens_nl} The temporal evolution of terms in the enstrophy transport equation for (a) $N=64$ and (b) $N=128$ test cases with no-limiter.}
   \end{center}
\end{figure}

\begin{figure}[htp!]
   \begin{center}
      \includegraphics[width=\textwidth]{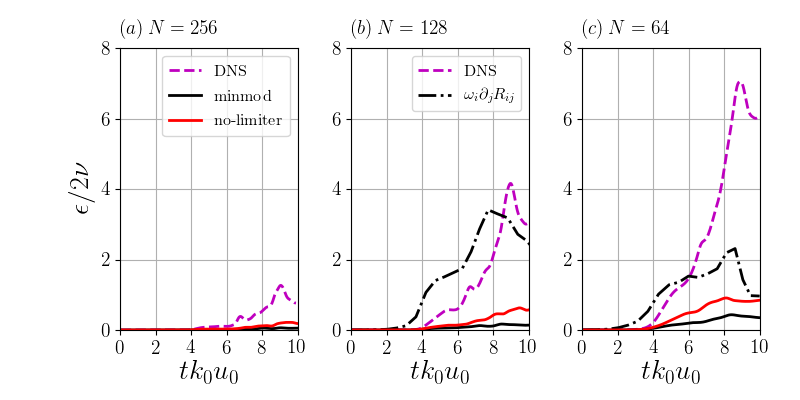}
      \caption{\label{fig:tg_ens_sgs} The SGS dissipation $\epsilon_{SGS}$ for the Taylor-Green vortex for (a) $N=256$, (b) $N=128$, and (c) $N=64$.  The SGS dissipation obtained from the modified equation analysis is shown for the $N=128$ and $N=64$.}
   \end{center}
\end{figure}

\begin{figure}
   \begin{center}
      \includegraphics[width=\textwidth]{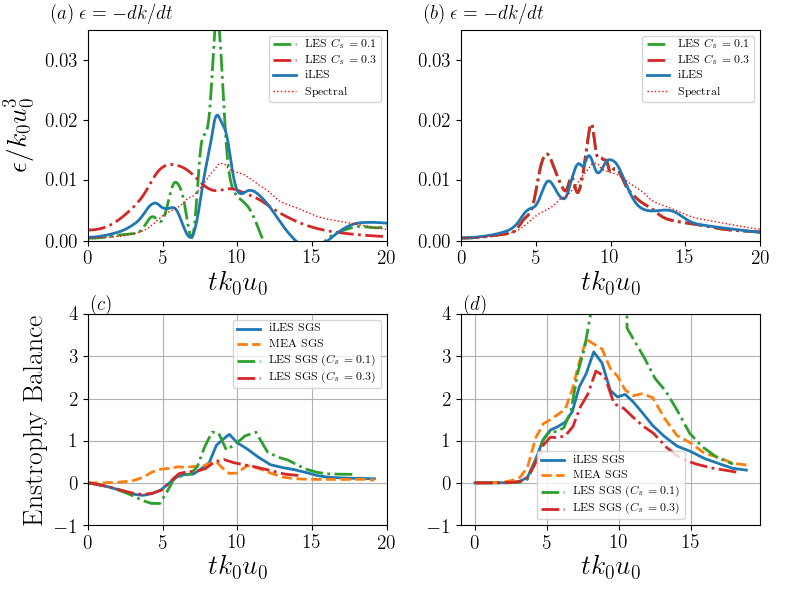}
      \caption{\label{fig:tg_les} The filtered dissipation $\widetilde{\epsilon}^*$  as a function of time for (a) $N=64$ and (b) $N=128$ comparing test case with minmod limiter, two explicit LES cases with $C_s=0.1$ and $0.3$, and DNS results from a spectral method.  The temporal evolution of the calculated SGS dissipation and diffusion obtained from the present method, the modified equation, and Smagorinsky model for explicit LES with grid discretization of (c) $N=64$ and (d) $N=128$.}
   \end{center}
\end{figure}

\begin{figure}
   \begin{center}
      \includegraphics[width=.8\textwidth]{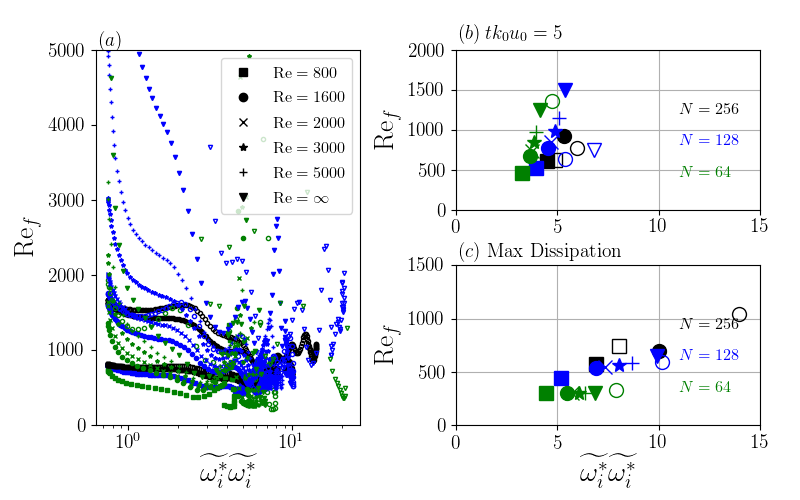}
      \caption{\label{fig:tg_ref}  The effective Re, $\textrm{Re}_f$, as a function of twice the enstrophy throughout the runtime of the simulations for several grid resolutions and viscosities.  The $\textrm{Re}_f$ at (b) $t=5$ and (c) maximum dissipation.}
   \end{center}
\end{figure}

\begin{figure}
   \begin{center}
      \includegraphics[width=\textwidth]{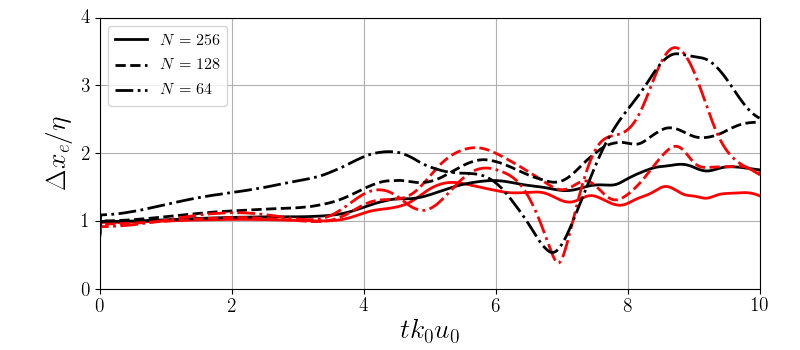}
      \caption{\label{fig:tg_dx}  The ratio of the effective grid spacing and Kolmorgorov lengthscale (from DNS) $\Delta x_f/\eta$ for the $\textrm{Re}=1600$ cases. Red: no-limiter, Black: minmod limiter. }
   \end{center}
\end{figure}

\indent The Taylor-Green vortex is a well-studied flow field \cite{frisch1995turbulence, brachet1991direct, brachet1983small}, which encompasses large structures, transition, and decaying turbulence, and is used to assess the ability of the scheme to capture the complexities of vortex stretching and breakdown using under-resolved grids \cite{zhou2014estimating, drikakis2007simulation, grinstein2011simulations}.  The test case is used to demonstrate how the present scheme can be employed to emulate the dominant sub-grid scales, most importantly dissipation, from the initial conditions of large coherent structures through a transition to turbulence at high $\textrm{Re}$.  
The initial velocity flow field for the Taylor-Green vortex is given by
\begin{align}
   \nonumber u_1(x_1,x_2,x_3) &= u_0 \cos(kx_1) \sin(kx_2) \cos(kx_3) \\
   u_2(x_1,x_2,x_3) &= -u_0 \sin(kx_1) \cos(kx_2) \cos(kx_3) \\
   \nonumber u_3(x_1,x_2,x_3) & =0
\end{align}
where $u_0=1$ is the velocity scale and $k_0=1$ is the length scale of the flow field.  Using these scales, the $\textrm{Re}= u_0 / \nu k_0 = 1 / \nu$, where $\nu$ is the dynamic viscosity specified for each simulations.  Simulations are designed in a periodic cube with length $2\pi$ with a uniform grid. We start by exploring how well the scheme captures the pertinent physics at $\textrm{Re} = 1600$ with a non-dissipative second-order central difference (referred to as no-limiter) and a high dissipative limiter, the minmod limiter.  The first limiter is directly related with the results from the modified equation in section \ref{sec:mea}. On the other hand, the modified equation for the first-order backward slope detailed in section \ref{sec:mea} is only similar to the minmod limiter because the minmod limiter chooses between first-order backward and forward differences and removes slopes where the backward and forward differences have opposite slopes.   \\
\indent Figure \ref{fig:tg_ee} shows the energy spectra of simulations at several grid resolutions ($N=64,128$ and $256$)  at two time instances $tu_0k_0=5$, during the transition to turbulence as the large-scale structures are breaking down, and $tu_0k_0=10$, near the maximum dissipation.  Additionally, a spectral method solver \cite{parish2017dynamic} is employed to provide direct numerical simulation (DNS) comparisons.  The DNS employs 256 spectral modes to fully resolve the flow.   
In Fig. \ref{fig:tg_ee}(a), the low wavenumber region of the energy spectra in all simulations is able to be captured accurately compared the DNS results from the spectral method regardless of the slope limiter employed.  The decline in energy in the spectrum for each simulation corresponds to the grid resolution where the lowest resolution begins to fall off at lowest wavenumbers.  This is expected for the under-resolved simulations.  The highest resolution case, while using the same number of degrees of freedom as the DNS, approaches the DNS solution but has slightly less energy in inertial range modes.  Test cases that employ the minmod limiter have lower energy in modes starting near a wavenumber $k/k_0=10$ compared to the no-limiter limiter. It becomes noticeable in the lower resolution test cases at lower wavenumbers, however, they all occur in the inertial range.  This is due to the dissipative nature of the minmod limiter, which allows for more implicit dissipation especially in modes that have a wavenumber comparable to the inverse size of the grid cell, $1/\Delta_x$.     
Figure \ref{fig:tg_ee}(b) shows similar behavior for the energy spectra; however, the lowest grid resolution cases show an increased energy at low wavenumbers.  This can be attributed to the evolution of the flow at this grid resolution because the number of necessary modes needed to provide an acceptable solution are higher than the number of modes resolved.  \\
\indent The Taylor-Green vortex is simulated as several higher $\mathrm{Re} = 2000, 3000, 5000$, and an inviscid case, $\mathrm{Re} = \infty$, in order to investigate the scheme in cases where the impact of viscosity is low or, in the case of $\mathrm{Re} = \infty$, nonexistent.  Figures \ref{fig:tg_ee_re}(a) and (b) show the energy spectra of each Re case at $tu_0k_0=5$ and $tu_0k_0=10$, respectively, at two different grid resolutions. All simulations show similar low and high wavenumber behavior for all Re at particular grid resolution.  All cases regardless of Re or grid resolution, the largest scales are similar.  In fact, the Reynolds number has little effect on the low wavenumbers, while at the high wavenumbers, the energy in these modes increase with Re.  This is due to the diminishing effects of the viscosity.  However, the spectra clearly show that the scheme provides a pathways for energy dissipation because there is no build up of energy in high wavenumber modes.    \\
%
\indent Next, we focus on the dissipation spectra of the simulations, which is directly obtained from the vorticity flow field.  Figure \ref{fig:tg_ens}(a) shows the dissipation spectra at $tu_0k_0=5$ has several low wavenumber peaks that are well captured by all simulations with $N=128$ or $256$ while there is lower dissipation in most modes for the $N=64$ simulations.  Overall, compared to the energy spectra, the dissipation spectra is well captured at all wavenumbers especially for the $N=256$ test cases.  This is an indication that these simulations can accurately resolve most of the dissipation.  Near the peak dissipation at $tu_0k_0=10$, Fig \ref{fig:tg_ens}(b) shows that grid resolution and the slope limiter start to have more effect on the resolved dissipation.  The $N=256$ no-limiter case dissipation spectrum is very comparable to the DNS dissipation spectrum at most wave numbers, while the minmod limiter test case at the same resolution reveals that it can only capture the dissipation spectra at lower wavenumbers.  A general trend shows that the no-limiter cases have slightly higher dissipation modes.  This is an indication that the no-limiter test cases can resolve slightly more dissipation compared to the minmod limiter, but we need to investigate further.  While the no-limiter can capture slightly more dissipation, it also does not add any numerical dissipation on the order of $\Delta_x^2$, so it may not be able to dissipate sufficient energy. The minmod case shows slightly lower dissipation modes in the spectra, but there is numerical dissipation that is not accounted for in this metric. \\
\indent The total dissipation, including numerical dissipation, $\epsilon = dK/dt$, where $K = \frac{1}{2}\langle u_i u_i \rangle$ ($\langle \cdot \rangle$ indicates averaged over the computational domain) is the kinetic energy, is integrated over the entire domain and recast into a non-dimensional form, $\epsilon^* = \epsilon / k_0 u_0^2$.   
Figure \ref{fig:tg_diss}(a) shows the temporal evolution of the dissipation for the $N=256$ test cases where both the no-limiter and minmod limiter cases provide reasonable results compared to the DNS dissipation, which is filtered using several sharp-cutoff spectral filters with widths including $\Delta = 2\Delta_x, 4\Delta_x$, and $8\Delta_x$. The filtered DNS dissipation $\widetilde{\epsilon}$ is the temporal derivative of the filtered kinetic energy $\widetilde{k}$. The filtered kinetic energy is obtained by integrating over all wavenumbers in the sharp-cutoff spectral filtered energy spectrum with different filter widths obtained from the DNS at each time step.  While there are some discrepancies the peak dissipation is captured reasonably well for both cases and is comparable to the DNS filtered solution at $\Delta = 2\Delta_x$.  
Figure \ref{fig:tg_diss}(b), which shows the total dissipation for the $N=128$ cases, reveals that the minmod limiter case out performs the no-limiter case.  The minmod limiter case can reasonably capture the peak dissipation, which is comparable to the $N=256$ cases and the filtered DNS dissipation at $\Delta = 4\Delta_x$.  Figure \ref{fig:tg_diss} (c) shows total dissipation for the $N=64$ test cases.  Neither results are as reasonable as the higher resolution cases indicating that very under-resolved cases may be overly dissipative in certain regimes such as near the maximum dissipation of the Taylor-Green vortex. On the other hand, the numerical dissipation in a regime of strong coherent vortex interaction at short times or decaying turbulence at long times are more reasonably captured in very under-resolved cases.  However, this low resolution case demonstrates evidence of SGS backscatter present in numerics as the numerical dissipation becomes negative around $tu_0k_0 = 15$. 
These results as well as the dissipation spectra indicate that when the flow is reasonably under-resolved, the numerical dissipation is necessary. The adequacy of the resolution will be explored below and is imperative to understand when employing ILES in general.  In our scheme, the numerical dissipation has a form that is qualitatively similar to physically devised models, an aspect that offers insight in understanding the implicit SGS dissipation mechanisms.\\  
\indent Simulations at several higher $\mathrm{Re} = 1600, 2000, 3000, 5000$, and $\infty$ reveal the effects of the numerical dissipation with diminishing viscosity for $N=128$ and $N=64$ in Figs. \ref{fig:tg_re_diss}(a) and (b), respectively.  For the $N=128$ between the initial condition and maximum dissipation ($0 < tu_0k_0 < 9$), all cases share a similar numerical dissipation. The extrema are amplified by the increasing Re, which suggests that viscosity plays a role in smooth the dissipation.  The effects at $\mathrm{Re}=\infty$ shows the largest discrepancies between the ILES cases and the DNS and suggest that including some physical viscosity dissipation helps stabilize the scheme.  Figure \ref{fig:tg_re_diss}(b) shows the $N=64$ cases with increasing Re.  The cases corroborate the findings in Fig. \ref{fig:tg_diss}(c) that a severely under-resolved case does not necessarily capture the dissipation well for all cases.  This will be further explored in further below.\\
\indent Next, we provide further analysis to compare the numerical dissipation of the scheme to the form of dissipation that is obtained through the modified equation analysis in section \ref{sec:mea}. Similar analysis \cite{cerutti2000spectral, da2004effect,zhou2020dependence}  on how the resolved vorticity field affects SGS dissipation transport has provided insights into the dynamics of the flow. We start with presenting the transport equation for the resolved enstrophy $\widetilde{\mathcal{E}} = \frac{1}{2} \langle \widetilde{\omega_i} \widetilde{\omega_i} \rangle$ that is obtained by multiplying Eqn. (\ref{eqn:vte_conserv_grp}) by $\omega_i$ and filtering.  The resolved enstrophy transport equation is given by the following:
\begin{nalign}
   \frac{\partial \widetilde{\mathcal{E}}}{\partial t} =
       &- \underbrace{\widetilde{u_j}\frac{\partial \widetilde{\mathcal{E}}}{\partial x_j}}_\text{I}
        + \underbrace{\vphantom{\frac{\partial}{\partial_j}}\widetilde{\omega}_i \widetilde{\omega}_j \widetilde{s}_{ij}}_\text{II}
        + \underbrace{\nu \frac{\partial^2 \widetilde{\mathcal{E}}}{\partial x_j \partial x_j}}_\text{III}
        - \underbrace{\nu \frac{\partial \widetilde{\omega_i}}{\partial x_j} \frac{\partial \widetilde{\omega_i}}{\partial x_j}}_\text{IV}  
       - \underbrace{\widetilde{\omega}_i  \widetilde{u}_i \frac{\partial \widetilde{\omega}_j}{\partial x_j} }_\text{V} 
        - \underbrace{\widetilde{\omega}_i \frac{\partial}{\partial x_j} \left ( R_{ij} \right )}_\text{VI},
   \label{eqn:resolved_ete}
\end{nalign}
which relates the temporal change in the resolved enstrophy to spatial changes in the vorticity, velocity, and SGS vorticity stress tensor.
The equation contains several mechanisms that balance the temporal change in resolved enstrophy: I.) convection, II.) amplification by vortex stretching, III.) diffusion by viscous effects, IV.) dissipation by viscous effects, V.) diffusion by divergence modification, and VI) SGS dissipation and SGS diffusion. Term VI can be expanded into two terms as follows:
\begin{equation}
    \widetilde{\omega}_i \frac{\partial}{\partial x_j} \left ( R_{ij} \right ) =  \frac{\partial }{\partial x_j} (\widetilde{\omega_i} R_{ij}) - R_{ij} \frac{\partial \widetilde{\omega_i}}{\partial x_j},
    \label{eqn:sgs_prod_expansion}
\end{equation}
where the first term is the diffusion by SGS modes and the second term is the SGS dissipation.  This expansion allows us to investigate the SGS dissipation by terms that are present in the modified equation.  We discretized each derivative (in space and time) using standard 3 point second order central differencing. Note that vorticity does not need to be discretized, because it is the solution variable.  Every term in Eqn. (\ref{eqn:resolved_ete}) except term VI can be directly calculated from the instantaneous flow.  However, because it is necessary to balance Eqn. (\ref{eqn:resolved_ete}), we can find term VI, which gives us the exact numerical SGS dissipation and SGS diffusion due to the scheme and slope limiter. Figure \ref{fig:tg_ens_bal_64}(a) shows the temporal evolution of each term in Eqn. (\ref{eqn:resolved_ete}) for the test case with $N=64$ employing the minmod limiter.  Near the initial time, the temporal change in the enstrophy is dominated by the amplification of vortex stretching.  At this time, the vortical structures are large and nearly resolved by the grid resolution and there is very little SGS interactions.  However, as the flow progresses towards the transition to turbulence around $tk_0u_0=5$, the SGS term begins to increase as a balance to the amplification by vortex stretching.  Around this time, there is non-negligible viscous dissipation, however, the other terms are small compared to the vortex stretching and SGS terms.  This trend continues towards the maximum dissipation time near $tk_0u_0=9$ and after as the turbulence begins to decay. The trend that vortex stretching balances the SGS dissipation is consistent with experiments of high Reynolds number turbulence \cite{cerutti2000spectral} with multi-probe hot-wire measurements.    Overall, the SGS term is balanced by the vortex stretching, which indicates that a majority of the dissipation is provided by the numerical method not the viscosity.  \\
\indent In Fig. \ref{fig:tg_ens_bal_64}(b) the SGS dissipation and SGS diffusion term (VI) in Eqn. (\ref{eqn:resolved_ete}) are compared with the modeled SGS dissipation and production (i.e. modified equation sub-grid dissipation and diffusion) that uses $R_{ij} = R^t_{ij} + R^g_{ij} + R^h_{ij}$ from Eqns. (\ref{eqn:mea1_t}), (\ref{eqn:mea1_grad}), and (\ref{eqn:mea1_hyper}), respectively.   The temporal evolution of the modified equation SGS dissipation and SGS diffusion has a similar trend as the numerically calculated term while there are some discrepancies near the peak dissipation.  These discrepancies arise from a) the simplified calculation of the modified equation by not incorporating the exact limiter definition, and b) truncation of the final modified equation to only second-order accurate terms.  The modeled SGS dissipation and modeled SGS diffusion are obtained through Eqn. (\ref{eqn:sgs_prod_expansion}), which shows that  SGS dissipation is the dominating term compared to the diffusion.  \\
\indent The balance of enstrophy is also investigated at a higher grid resolution in Fig. \ref{fig:tg_ens_bal_128}(a) with $N=128$ and the minmod limiter.  The temporal evolution of the amplification by vortex stretching term is the dominant term in the enstrophy balance.  Compared the $N=64$, the amount of the viscous dissipation is increased, which is expected because the higher grid resolution resolves more viscous dissipation.  The increased viscous dissipation acts as an offset for the SGS dissipation and SGS diffusion term, which is comparably less for all simulated time compared to the corresponding terms in the $N=64$ test case.   Figure \ref{fig:tg_ens_bal_128}(b) shows the comparison of the calculated SGS dissipation and diffusion term with the modeled SGS dissipation and SGS diffusion.  At this resolution, the model elucidated from the modified equation analysis is shown to be well represented because higher order terms in the modified equation become smaller with higher grid resolution.  As with the coarser grid resolution the SGS dissipation is dominant compared to the SGS diffusion.  Overall, the enstrophy transport equation analysis shows that the numerical results and the form of the implicit SGS model derived from the modified equation analysis are similar and provides validation for the modified equation analysis for this scheme.  \\
%
\indent The modified equation SGS dissipation and diffusion can be separated into three different terms based on Eqns. (\ref{eqn:mea1_grad}), (\ref{eqn:mea1_hyper}) and (\ref{eqn:mea1_t}) developed directly from the modified equation.  This allows us to identify which mechanism dominants the dissipation in the ILES.  Figure \ref{fig:tg_ind_ens}(a) shows the temporal evolution of the three terms for the $N=64$ test case with the minmod limiter.  Overall the modified equation SGS dissipation is driven by terms similar to a tensor-gradient model.  Similarly, for the $N=128$ test case with the minmod limiter shown in Fig. \ref{fig:tg_ind_ens}(b), the tensor-gradient terms in Eqn. (\ref{eqn:mea1_grad}) are the dominant mechanism for dissipation.  In general, these SGS models with tensor-gradient are not overly dissipative and normally are paired with other models.  This is consistent with the SGS dissipation shown herein.  The modified equation terms similar to a hyper-viscosity model in Eqn. (\ref{eqn:mea1_hyper}) are shown to contribute to the total SGS dissipation, which enables the total modified equation SGS dissipation to have a similar temporal evolution as the calculated SGS dissipation (term VI).  The SGS dissipation contribution from Eqn. (\ref{eqn:mea1_t}) is shown to relatively less significant for both resolutions.  \\ 
\indent Figure \ref{fig:tg_ens_nl}(a) shows the enstrophy balance for the $N=64$ with no-limiter case to show the temporal evolution of the transport of enstrophy when no limiter is employed in the scheme, which limits the numerical dissipation.  Similar to the limited cases presented in Fig. \ref{fig:tg_ens_bal_64}, amplification by vortex stretching dominants the evolution of enstrophy, but is significantly larger than the limited cases.  Furthermore, terms IV and V are also augmented compared to the limited case by the lack of limiter and pathway for dissipation.  This has a large influence on the balance and the SGS dissipation/diffusion term.  This term is significantly higher than than the limited case and corroborates the high numerical dissipation observed in Fig. \ref{fig:tg_diss}(c).  Similarly, the $N=128$ case shows that the amplification of vortex stretching is elevated by the lack of limiting.  This affects the enstrophy balance by increasing the remainder calculated in the SGS dissipation/diffusion.     \\
Figure \ref{fig:tg_ens_sgs} compares the SGS dissipation $\epsilon_{SGS}$ with the expected SGS dissipation obtained through DNS. The latter is obtained by filtering the dissipation spectra Fig. \ref{fig:tg_ens} using a sharp-cutoff spectral filter with a length scale $2\pi / (N/4+1)$, where $N$ is the grid resolution for each case.  Figures \ref{fig:tg_ens_sgs}(a), (b), and (c) show that the SGS dissipation for the $N=256, 128$, and $64$ cases, respectively.    As expected, as the length scale of the filter is increased, the total SGS dissipation increases for both the DNS and ILES. 
The SGS dissipation from both ILES cases (minmod and no-limiter) are lower than the DNS dissipation as expected based on the dissipation spectra, which indicated that the total resolved dissipation smaller than the filter scale is relatively small.
The SGS dissipation obtained through the modified equation analysis for the $N=128$ and $N=64$ cases with the minmod limiter $\omega_i \partial_j R_{ij}$ is included in Figs. \ref{fig:tg_ens_sgs}(b) and (c).  The SGS dissipation is relatively similar to that from the filtered DNS for $N=128$. It is however,  under-predicted for $N=64$ .  This provides further evidence with the Fig. \ref{fig:tg_diss}(c) that more analysis and clarification is needed especially in the low resolution limit.  \\
%
\indent Next, to further understand the implicit numerical dissipation and SGS model derived from the modified equation analysis we compare several terms of the modified equation to the well-known Smagorinsky model.  
A Smagorinsky model \cite{smagorinsky1963general} was developed in Ref. \cite{mansfield1998dynamic} for the VTE and can used for closure of $R_{ij}$:
\begin{equation}
\frac{\partial R_{ij}}{\partial x_j} = g_i = - \frac{\partial}{\partial x_j} \left ( \nu_{t} \frac{\partial \omega_i}{\partial x_j} \right ) - \frac{\partial \nu_t}{\partial x_j} \frac{\partial \omega_j}{\partial x_i}.
\label{eqn:LES_subgrid_eq}
\end{equation}
The eddy viscosity $\nu_{t}$ is given by
\begin{equation}
\nu_{t}= C_{s}{\Delta}^{2}|\widetilde{S}|,
\label{eqn:LES_eddyviscosity_eq}
\end{equation}
where $\widetilde{S}_{ij}$ is the filtered strain-rate tensor, $C_{s}$ is the Smagorinsky constant and $|\widetilde{S}|= (2\widetilde{S}_{ij}\,\widetilde{S}_{ij})^{\frac{1}{2}}$. While the implicit model is shown by modified equation analysis to consist of several terms (Eqns. (\ref{eqn:mea1_t}), (\ref{eqn:mea1_grad}), and (\ref{eqn:mea1_hyper})) and provide implicit SGS vorticity stress on all tensor element, the comparable Smagorinsky model creates an anti-symmetric SGS vorticity stress tensor. \\
\indent Additional LES cases are simulated using the explicit model in Eqn. (\ref{eqn:LES_subgrid_eq}). No limiter is used in the upwind based scheme to ensure that the dissipation is enabled through the explicit model; however, there is still interaction between the scheme dissipation and explicit model dissipation.  Several Smagorinsky constants are employed including $C_s = 0.1$ and $0.3$, where the former is used in Ref. \cite{mansfield1998dynamic} with a vortex particle method discretization.  Simulations are performed at $N=64$ and $128$ resolution of the Taylor-Green vortex at $\mathrm{Re} =1600$ to compare to the present results.  Figures \ref{fig:tg_les}(a) and (b) show the numerical filtered dissipation $\widetilde{\epsilon}^*$ at $N=64$ and $N=128$, respectively.  The dissipation is shown to be affected significantly by the Smagorinsky constant.  In fact, the constant used previous studies is shown to be out performed by the ILES.  The higher constant, especially in $N=64$ performs slightly better.   The discrepancies are affected by the interactions of the modified equation of the vorticity transport scheme, even if no limiter is employed, and the SGS model. Further insights into the performance of explicit LES compared to the present implicit model is shown in the enstrophy transport analysis in Figs. \ref{fig:tg_les}(c) and (d) for the $N=64$ and $N=128$, respectively.   Results from the explicit LES model suggest that the value of the model coefficient produces high variability for the SGS dissipation. Moreover, \emph{a priori} tests of forced homogeneous, isotropic turbulence~\cite{mansfield1998dynamic} reveal that both vorticity convection and vortex stretching contribute to sub-grid scale dissipation, but the explicit eddy-viscosity model does not capture the contribution from vortex stretch adequately. While the variability may be reduced with a dynamic model~\cite{mansfield1998dynamic,germano1991dynamic}, the overall simplification, deterministic dissipation through the modified equation, consistent integration both the vorticity convection and vortex stretching, and reduced computational effort of the implicit approach are desirable qualities for a scheme.  Furthermore, ILES has been  described by one constant---typically the effective viscosity~\cite{zhou2016comparison}---to quantify the observed behavior of the simulated flow field.      \\
\indent Finally, we analyze the scheme by attempting to provide methodology to characterize the flow field in terms of a numerical effective Reynolds number $\textrm{Re}_f$.  In the previous analysis, we demonstrated that the numerical dissipation of the present schemes and the dissipation predicted by the modified equation are similar.  In what follows, we provide methodology to characterize the flow in order to determine if the simulation is sufficiently resolved to provide an accurate solution.  This is particularly important considering the $N=64$ simulations where numerical dissipation is considerably different from DNS dissipation. If the simulation is not sufficiently resolved then we provide information on what the simulation physically solved.  In coarse-grained simulations, such as LES either explicit or implicit, where small scales are not directly solved but are modeled, the $\textrm{Re}_f$ can be used to interpret what the simulation physically solved.  The methodology is based on the analysis of an ILES Euler scheme of turbulence flows by Ref. \cite{zhou2014estimating}.
For incompressible flows with sufficiently fine grid resolution to be considered a DNS, the numerical dissipation $-dK/dt$ and resolved dissipation $\epsilon_\mathcal{E} = 2 \nu \langle \omega_i \omega_i \rangle$ are equal. This means that the scales impacted by the numerics of scheme are in only in a narrow band of high wavenumbers dominated by the viscous scales.   In the under-resolved simulations presented herein, the ratio of the resolved enstrophy and dissipation is used to characterize the $\textrm{Re}_f$ as the following:
\begin{equation}
   \textrm{Re}_{f} = \frac{u_0}{k_0 \nu_f} = \frac{\langle \omega^*_i \omega^*_i \rangle}{\epsilon^*}.
   \label{eqn:ref}
\end{equation}
\indent Figure \ref{fig:tg_ref}(a) shows the temporal evolution of $\textrm{Re}_{f}$ of several test cases including some additional test cases with different $\textrm{Re} = u_0/k_0 \nu$ than presented above in order to understand some general trends. The additional Re included are 800, 2000, 3000, 5000, and an inviscid case where the $\textrm{Re} = \infty$. 
In each case, the initial $\textrm{Re}_{f}$ is near the $\textrm{Re}$ indicating that the initial numerical dissipation and resolved dissipation are equal.  However, the $\textrm{Re}_{f}$ decreases as the enstrophy increases towards a maximum.  The grid resolution of each case has an effect on how the $\textrm{Re}_{f}$ decreases where the largest decreases can be attributed to the lowest grid resolution cases.   
In Fig. \ref{fig:tg_ref}(b), the $\textrm{Re}_{f}$ is selected at $t=5$ when the transition to turbulence is occurring.  The $\textrm{Re}_{f}$ is observed to be a function of the physical viscosity while the grid resolution has a secondary effect.  However, the $\textrm{Re}_{f}$ lower than the Re, which indicates there is noticeable discrepancy in the numerical dissipation and the resolved dissipation.   
The $\textrm{Re}_{f}$ is selected at the location of maximum dissipation for each case and is shown in Fig. \ref{fig:tg_ref}(c).  The results show that the grid resolution has the largest effect on $\textrm{Re}_{f}$.   The $\textrm{Re}_{f}$ increases with the grid resolution.  Regardless of the slope limiter or $\textrm{Re}$, all cases show similar behavior.  This suggests that the numerical dissipation is highly affected by the grid resolution compared to the effects of the viscous diffusion or limiting in the flux functions of the present scheme for the Taylor Green vortex.  
We observe that the $N=128$ cases, which have a closer $\textrm{Re}_{f}$ to 800 than 1600, can be reasonably compared to the results for the $\textrm{Re} = 800$ in some parts of the evolution of the flow. 
At the highest resolution case $N=256$, secondary effects of the $\textrm{Re}$ begin to have a larger effect.    \\
\indent Next, we can use the ratio of the numerical dissipation and the resolved dissipation to create an effective length scale as follows:
\begin{equation}
   \Delta x_f = \frac{\epsilon^{1/2}}{ \left ( \widetilde{\omega_i} \widetilde{\omega_i} \right )^{3/4}}.
   \label{eqn:dxf}
\end{equation}
The effective length scale is compared with the Kolmogorov length scale $\eta$ provide by the DNS at a $\textrm{Re}=1600$ in Fig. \ref{fig:tg_dx}.  The temporal evolution of the effective length scale indicates that before the transition to turbulence, the flow field is relatively resolved.  However, as the large structures begin to break down, the grid is no longer sufficiently fine to resolve the flow and the effective length scale become larger than the Kolmogorov scale.  The grid resolution has a large effect on the evolution of the effective length scale. \\
\indent Further analysis into $\textrm{Re}_{f}$ of the simulations allows us to interpret the effective viscosity $\nu_f$ with the Taylor micro-scale $\lambda$ which is calculated as follows:
\begin{equation}
   \lambda = \frac{1}{3} \sum^3_i \sqrt{\langle u_i u_i \rangle / \langle \frac{\partial u_i}{\partial x_i}\frac{\partial u_i}{\partial x_i} \rangle}.
   \label{eqn:lambda}
\end{equation}
The effective Taylor micro-scale $\textrm{Re}_\lambda = u^\prime \lambda/\nu_f$.  Many studies on high-Re turbulence have investigated the relationship between the energy containing eddies in flows with the dissipation especially using turbulence-resolving DNS or experiments \cite{kaneda2003energy, falkovich1994bottleneck, wang1996examination, sreenivasan1998update}.  From these efforts, we are able to conclude that the time scale of the energy-containing eddies $u^\prime/L$ are on the same magnitude as the time scale of the dissipation rate $\epsilon/u^{\prime2}$, where $u^\prime$ is the characteristic velocity scale and $L$ is the characteristic length scale.  This key finding is especially beneficial for success with under-resolved simulations where by resolving large energy-containing eddies, there is some understanding of the magnitude of the dissipation just based on resolved characteristics.  Furthermore, the DNS studies showed that with a sufficiently high $\text{Re}_\lambda$, the flow reaches a minimum state \cite{zhou2007unification} where turbulence seems to exhibit self-similar behavior.  The minimum state of turbulence flows where the time scale of the large scales and dissipation in terms of $\textrm{Re}_\lambda$ can be as low as $\textrm{Re}_\lambda > 100$ \cite{sreenivasan1984scaling,dimotakis2000mixing} or  $\textrm{Re}_\lambda > 200$ \cite{kaneda2003energy, sreenivasan1998update}.  For the Taylor-Green vortex case with $\textrm{Re}=1600$, the maximum $\textrm{Re}_\lambda = 140$~\cite{brachet1991direct}. This implies that in order to reach some self-similarity of small-scale turbulence the flow must have sufficient separate of scales, and the dissipation must occurs away from the large scales in the spectral sense.  Ref. \cite{zhou2014estimating} used this criteria in order to characterize ILES schemes for velocity-pressure formulation to see if a minimum state is reached. Furthermore, well-known relationships for $\textrm{Re}_\lambda$ to a large-scale $\textrm{Re}_L$ exist. Several studies have pointed towards $\textrm{Re}_L \sim \textrm{Re}^2_\lambda$ \cite{tennekes1972first, sagaut2008homogeneous}.  \\   
\indent We present details on the characteristics of several of the Taylor-Green simulations in Table \ref{tbl:tg}, which shows the effective Re at transition to turbulence ($t=5$), maximum dissipation ($t\approx9$), and decaying turbulence ($t=15,20$).  Here, we calculate the characteristic velocity as the rms (root-mean-square) of the velocity fluctuations $u^\prime = \langle (\frac{1}{3} u_i u_i)^{1/2}\rangle$.  The high-resolution cases with $N=256$ show that the limiter has minor effects on $\textrm{Re}_f$ as also shown in Fig. \ref{fig:tg_ref}.  While only at the transition stage has the flow field reached a sufficiently high $\textrm{Re}_\lambda$, this is consistent with ILES simulations with for the Euler equations \cite{zhou2014estimating}.  The DNS has an initial $\mathrm{Re}_\lambda = 55$ which increases to  the maximum $\textrm{Re}_\lambda = 140$~\cite{brachet1991direct,zhou2016comparison} around $t=9$. $\mathrm{Re}_\lambda \approx 100$ at larger times $t>10$ as the turbulence decays.  In the $N=128$ test cases, the numerical dissipation provided by the minmod limiter has more effects while in the $N=64$, the $\textrm{Re}_f$ remains relatively low and does not provide positive results during turbulence decay indicating that the grid resolution is not high enough or the dissipation is not enough for this flow.     
\begin{table}[]
\centering
\caption{Effective Re for ILES of Taylor-Green Vortex}
\label{tbl:tg}
\begin{tabular}{lcrr@{\hskip 0.6cm}crr}
\hline
$N$ & Limiter    & \multicolumn{1}{c}{$t$} & $Re_f = \frac{u_0}{k_0 \nu_f}$ & $\Delta x_f$ & $Re_\lambda = \frac{u^\prime \lambda}{\nu_f}$ & $Re_L \sim Re_\lambda^2$ \\ \hline\hline
256 & No-limiter &  5.00 & 774  &  2.30$\times 10^{-2}$ & 98  & 9692     \\
    &            &  9.03 & 1048 &  1.60$\times 10^{-2}$ & 58  & 3407     \\
    &            & 15.00 & 1624 & 1.61$\times 10^{-2}$ & 47 & 2251       \\
    &            & 20.00 & 1704 & 1.98$\times 10^{-2}$ & 43 & 1871       \\ 
    & Minmod     &  5.00 & 923  &  2.16$\times 10^{-2}$ & 122 & 14991    \\
    &            &  8.30 & 698  &  2.13$\times 10^{-2}$ & 52  & 2711     \\
    &            & 15.00 & 1135 & 2.08$\times 10^{-2}$ & 47 & 2226       \\
    &            & 20.00 & 1283 & 2.30$\times 10^{-2}$ & 45 & 2048       \\ \hline
128 & No-limiter &  5.00 & 639  &  2.60$\times 10^{-2}$ & 86  & 7527     \\
    &            &  8.49 & 589  &  2.31$\times 10^{-2}$ & 42  & 1819     \\
    &            &  15.00 & 1328 & 1.92$\times 10^{-2}$ & 45 & 2100      \\
    &            &  20.00 & 1596 & 2.04$\times 10^{-2}$ & 41 & 1725      \\
    & Minmod     &  5.00 & 769  &  2.47$\times 10^{-2}$ & 109 & 11956    \\
    &            &  8.29 & 533  &  2.67$\times 10^{-2}$ & 52  & 2755     \\
    &            &  15.00 & 856 & 2.57$\times 10^{-2}$ & 39 & 1559       \\
    &            &  20.00 & 1313 & 2.39$\times 10^{-2}$ & 47 & 2212      \\ \hline
64  & No-limiter &  5.00 & 1356 &  1.84$\times 10^{-2}$ & 203 & 41296    \\
    &            &  8.23 & 329  &  3.28$\times 10^{-2}$ & 38  & 1503     \\
    & Minmod     &  5.00 & 667  &  2.80$\times 10^{-2}$ & 104 & 10965    \\
    &            &  8.33 & 303  &  3.75$\times 10^{-2}$ &  46 & 2144     \\
\hline 
\end{tabular}
\end{table}
\subsection{Forced Isotropic Turbulence}\label{sec:forced}
\begin{figure}
   \begin{center}
      \includegraphics[width=.95\textwidth]{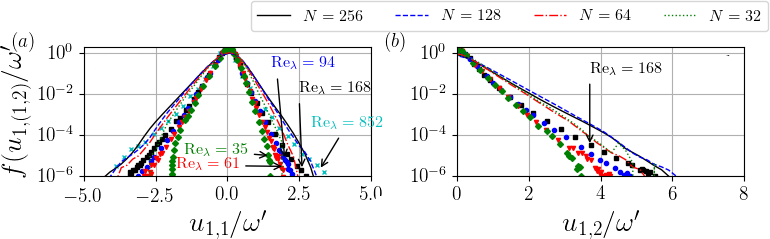}
      \caption{\label{fig:forced_pdf1} Longitudinal (left) and transverse (right) velocity gradient probability density functions normalized by the rms vorticity $\omega^\prime$ 
with several grid resolutions.  Markers indicate PDFs captured from DNS results from Ref. \cite{jimenez1993structure} are at $\textrm{Re}_\lambda=35, 61, 94$ and $168$ and Experimental measurements from Ref. \cite{castaing1990velocity} are at $\textrm{Re}_\lambda=852$. Each PDF is labeled.}
   \end{center}
\end{figure}

\begin{figure}
   \begin{center}
      \includegraphics[width=\textwidth]{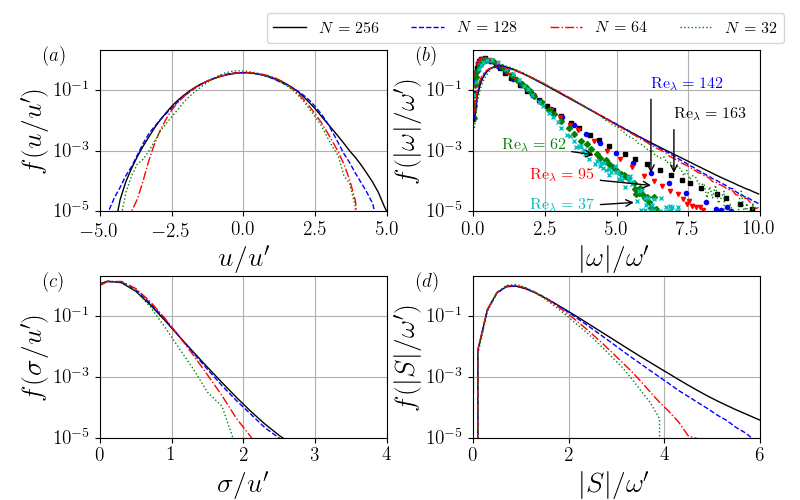}
      \caption{\label{fig:forced_pdf2} (a) Velocity magnitude, (b) vorticity magnitude, (c) vortex stretching, and (d) strain rate magnitude probability density function normalized by the rms vorticity $\omega^\prime$ of forcing case for several grid resolutions. The case $N=256$ with no-limiter is shown as a solid cyan line.  Marked indicate DNS results from Ref. \cite{jimenez1998characteristics} are at $\textrm{Re}_\lambda=37, 62, 95, 142$ and $168$. Each PDF is labeled.}
   \end{center}
\end{figure}

\begin{figure}
   \begin{center}
      \includegraphics[width=\textwidth]{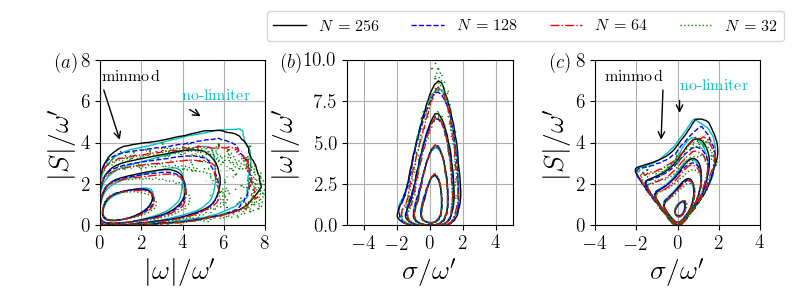}
      \caption{\label{fig:forced_jpdf} Turbulence statistics of the forced turbulence simulations 
for grid resolutions of $N=256, 128, 64$ and $32$.  Contour lines for each plot are shown at $\{10^0, 10^{-1}, 10^{-2}, 10^{-3}\}$.}
   \end{center}
\end{figure}

\begin{figure}
   \begin{center}
      \includegraphics[width=\textwidth]{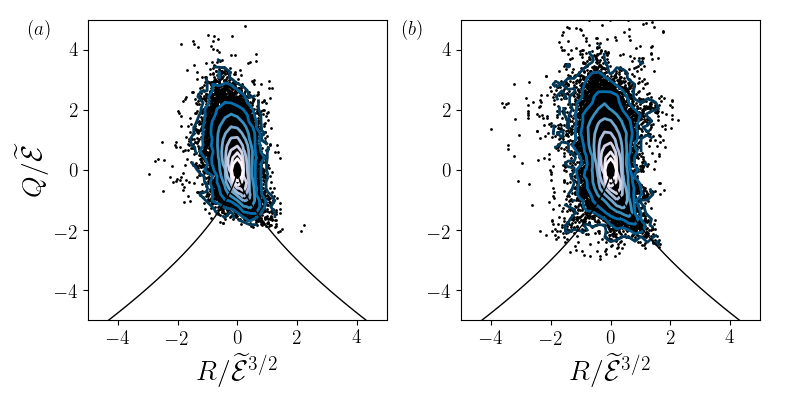}
      \caption{\label{fig:forced_qr} Joint PDF of velocity gradient invariants $Q$ and $R$ normalized by the resolved enstrophy $\widetilde{\mathcal{E}}$ of the forced turbulence simulations 
for grid resolutions of (a) $N=64$ and (b) $N=128$.  Contour lines for each plot are shown in log-space between $0$ and $10^{-5}$.}
   \end{center}
\end{figure}

\begin{figure}
   \begin{center}
      \includegraphics[width=.95\textwidth]{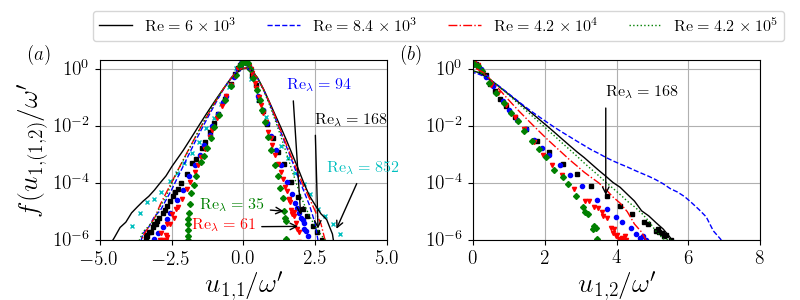}
      \caption{\label{fig:forced_re1} Longitudinal (left) and transverse (right) velocity gradient probability density functions normalized by the rms vorticity $\omega^\prime$ 
for grid resolutions of $N=64$ at several initial Re numbers.  Markers indicate DNS results from Ref. \cite{jimenez1993structure} are at $\textrm{Re}_\lambda=35, 61, 94$ and $168$. Experimental measurements from Ref. \cite{castaing1990velocity} are at $\textrm{Re}_\lambda=852$. Each PDF is labeled.}
   \end{center}
\end{figure}

\begin{figure}
   \begin{center}
      \includegraphics[width=\textwidth]{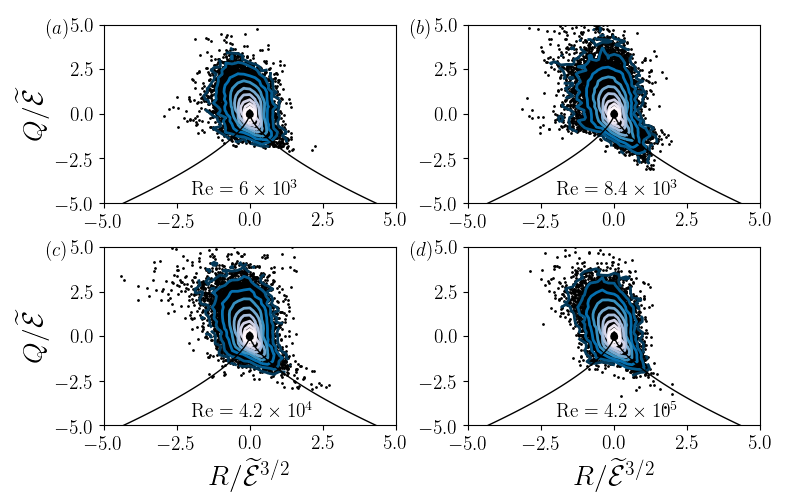}
      \caption{\label{fig:forced_qr_re} Joint PDF of velocity gradient invariants $Q$ and $R$ normalized by the resolved enstrophy $\widetilde{\mathcal{E}}$ of the forced turbulence simulations 
for grid resolutions of $N=64$ at several Re numbers.  Contour lines for each plot are shown in log-space between $0$ and $10^{-5}$.}
   \end{center}
\end{figure}

\begin{figure}
   \begin{center}
      \includegraphics[width=\textwidth]{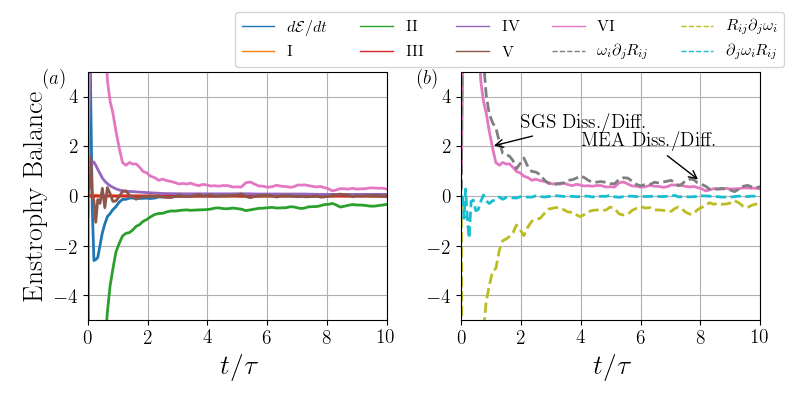}
      \caption{ \label{fig:forced_ens_bal_64} (a) The temporal evolution of terms in the enstrophy transport equation for the $N=64$ test case with minmod limiter. (b) The temporal evolution of the calculated SGS dissipation and diffusion and the temporal evolution of the SGS transfer and diffusion obtained from the modified equation for the $N=128$ test case with minmod limiter. }
   \end{center}
\end{figure}

\begin{figure}
   \begin{center}
      \includegraphics[width=\textwidth]{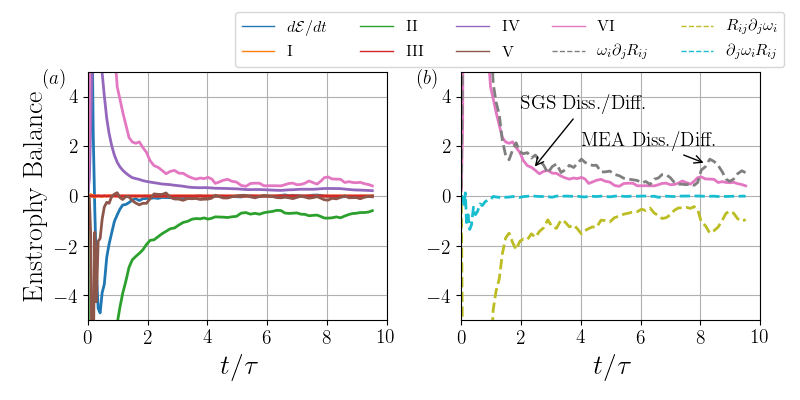}
      \caption{ \label{fig:forced_ens_bal_128} (a) The temporal evolution of terms in the enstrophy transport equation for the $N=128$ test case with minmod limiter. (b) The temporal evolution of the calculated SGS dissipation and diffusion and the temporal evolution of the SGS transfer and diffusion obtained from the modified equation for the $N=128$ test case with minmod limiter. }
   \end{center}
\end{figure}

\begin{figure}
   \begin{center}
      \includegraphics[width=.75\textwidth]{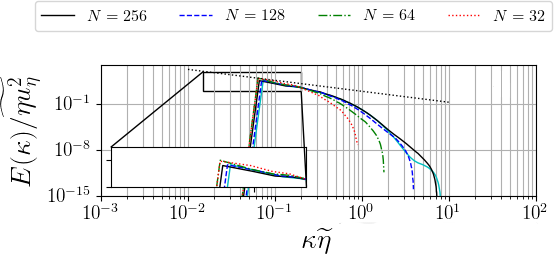}
      \caption{ \label{fig:forced_spec} Energy spectra of forced turbulence simulations. 
The thin dotted line is at a $-5/3$ slope. The case $N=256$ with no-limiter 
is shown as a solid cyan line.  }
   \end{center}
\end{figure}

\begin{figure}
   \begin{center}
      \includegraphics[width=.75\textwidth]{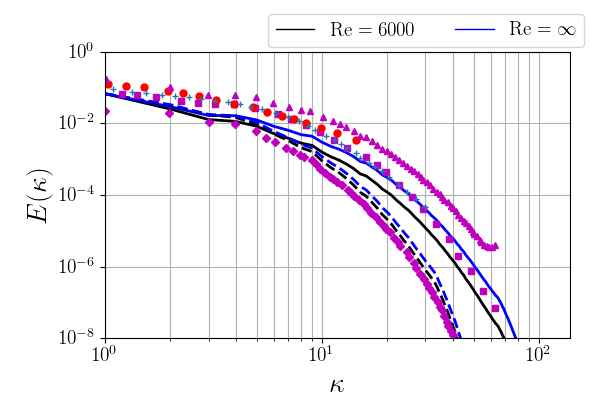}
      \caption{ \label{fig:forced_spec_re} Energy spectra of forced turbulence simulations at $\mathrm{Re} = 6000$ and $\mathrm{Re} = \infty$.  The solid line is $N=128$ and the dashed line is $N=64$. The diamond, square, and triangle markers are forced turbulence cases 1, 2, and 3, respectively, from Ref. \cite{rosales2005linear}. Circle and plus markers denote $N=64$ and $N=128$, respectively, and correspond to explicit LES of case 2 from Ref. \cite{piomelli2015grid}.}
   \end{center}
\end{figure}

\indent Next, we consider the numerical experimentation of isotropic turbulence simulation employing a forcing scheme developed for linear forced turbulence in physical space~\cite{rosales2005linear}.  Forced isotropic turbulence provides test cases that allow for velocity and vorticity statistics to be collected over the simulation, which are used to assess the ILES dissipation and ability to capture turbulence statistics.
A forcing term $\widetilde{\bm{f}} = \nabla \times \epsilon_0/3 u^{\prime2} \bm{u}$, where $\epsilon_0$ is a constant energy injection and $u^{\prime} = \left ( \frac{3}{2} \langle K \rangle \right )^{1/2}$ is the rms (root-mean-squared) velocity updated throughout the simulation, is included in Eqn. (\ref{eqn:vte_conserv_grp}) to maintain a nearly constant energy after an initial transient time. 
The energy injection constant is $\epsilon_0 = 5 \times 10^{-5}$.
The simulation cases are initialized using the conditions for isotropic turbulence.  It is simulated from an initial condition, which quickly decays into fully-developed turbulence, given in Refs. \cite{mansour1994decay} and \cite{rogallo1981numerical} where the initial energy spectrum takes the form:
\begin{equation}
   E(k,0) = \frac{3}{2A} \frac{1}{k_p^{\sigma+1}} k^\sigma \exp \left ( - \frac{\sigma}{2} \left ( \frac{k}{k_p} \right )^2 \right ),
\end{equation}
where $k_p$ is the wavenumber at which $E(k,0)$ is maximum,  $\sigma$ is parameter, and $A$ is $\int_0^\infty k^\sigma \exp (- \sigma k^2/2) dk$.  A flow is initialized with $\sigma = 4$ and $k_p = 3$. The $\nu = 0.0007$ gives a initial Reynolds number based on the length of the side domain, $2\pi$, and the initial velocity fluctuations of $\textrm{Re} = 6\times 10^3$.  Grid resolutions are employed in the $2\pi$-sided periodic box with the number of grid cells per dimension $N=256, 128, 64$, and $32$ with a minmod limiter for all resolutions.  An additional no-limiter case is employed for the $N=256$ resolution based on results above which show that the $N=256$ is nearly resolved for the present initialization.\\
%
%
\indent Several studies \cite{zhou2014estimating, wachtor2013implicit, fureby2002large} have employed forced turbulence simulations to examine the turbulence statistics especially to see sensitivities the velocity, vorticity, and strain rates have for ILES with different grid resolutions.
It has been observed in Ref. \cite{zhou2014estimating} for ILES that above a mixing transition of $\textrm{Re}_\lambda > 100$ \cite{dimotakis2000mixing}, the behavior of the turbulence statistics in terms of velocity probability density functions (PDFs) begin to  approach asymptotic Re statistics.  Figure \ref{fig:forced_pdf1} shows the longitudinal and transverse fluctuating velocity gradients with a comparison to DNS \cite{jimenez1993structure} and grid turbulence experimental measurements \cite{castaing1990velocity}.  The present simulations capture the turbulence statistics, especially, the high probability statistics well for all grid resolutions.  The tails of the PDFs, which represent only a small fraction of the solution correspond well with higher $\textrm{Re}_\lambda$ DNS and experimental cases, which suggests that the numerical dissipation in the scheme may be appropriate for simulating high Re turbulence cases in which intermittency and non-Gaussian behavior at the tail of PDFs are observed. 
The negative bias of the longitudinal derivative of the fluctuating velocity is similar to results present in DNS and experimental flow field statistics. 
The PDF of the transverse derivative of the fluctuating velocity shown in Fig. \ref{fig:forced_pdf1} demonstrates the convergence of the velocity towards non-Gaussian behavior observed for this quantity.   The turbulence statistics indicate that there is convergence towards high Re turbulence statistics. \\  
\indent Figure \ref{fig:forced_pdf2}(a) show the near Gaussian behavior, similar to flow fields from DNS results, of the velocity fluctuations, which show asymptotic convergence as the grid resolution increases. This also suggests that the effective Re is increasing with the grid resolution as seen in the Taylor-Green vortex.  
The statistics for the magnitude of the vorticity fluctuations are shown in Fig. \ref{fig:forced_pdf2}(b), which suggests that vorticity statistics are better represented compared to the velocity statistics at lower grid resolutions and for this particular Re.  Interestingly, this is not seen ILES of velocity-pressure formulations \cite{zhou2014estimating}, which may suggest that the present scheme designed to preserve vorticity may enhance the ability of lower grid resolution cases to capture some vorticity statistics more efficiently and  DNS simulations from Ref. \cite{jimenez1998characteristics} have relatively low $\textrm{Re}_\lambda$.  
Next, we focus on statistics related to the resolved enstrophy transport equation shown in Eqn. (\ref{eqn:resolved_ete}).  The PDF of the vortex stretching, $\sigma_{ij} = \omega_i s_{ij} \omega_j / |\omega|^2$, shows the convergence of the high probability portion of the PDF and tails that correspond well with different Re.   
Furthermore, the PDF of the strain rate magnitude in Fig. \ref{fig:forced_pdf2}(d) shows similar behavior.  Overall, the velocity and vorticity statistics display that the turbulence characteristics have asymptotic convergence and tails of the PDF resemble higher Re statistics with increasing grid resolution.  \\
\indent The effect of numerics of the ILES on the correlation of turbulence statistics is shown in Fig. \ref{fig:forced_jpdf}.  
First, Fig. \ref{fig:forced_jpdf}(a) shows that the joint PDF of the vorticity magnitude and the strain rate magnitude $|S|$. The correlation between the two quantities reveals some convergence towards in the tails of the PDF for the three lowest grid resolutions.  However, because the $N=256$ cases employs a central difference for the limiter, due to it being nearly resolved, there is a difference between this case and the other three.  Furthermore the correlations between the vorticity and vortex stretching and the strain rate magnitude and stretching in Figs. \ref{fig:forced_jpdf}(b) and (c) respectively show similar behavior.  The statistics indicate that there is high probability that the grid resolution does not have an effect on the statistics of the flow field, while the outlying statistics have some dependence on the resolution of the simulation. \\
\indent In order to further assess the implicit method, invariants of the velocity gradient~\cite{cantwell1993behavior}, which is related to turbulence dissipation, are analyzed. The three invariants of the velocity gradient $A_{ij} = \frac{\partial u_i}{\partial x_j}$ are given as follows:
\begin{nalign}
    P &= -A_{ii} = 0 \\
    R &= -\frac{1}{2} A_{ij} A_{ji} \\
    Q &= -\frac{1}{3} A_{ij} A_{jk} A_{ki}
\end{nalign}
Because the first invariant $P$ is zero for an incompressible flow, the velocity gradient is completely determined by the second and third invariants, $R$ and $Q$, respectively.  The invariants determine the local flow topology and indicate stretching and stability. \\
\indent Figure \ref{fig:forced_qr}(a) and (b) shows the joint PDF for the $N=64$ and $N=128$ test cases, respectively.  The invariants are determined at every grid points at $t/\tau = 10$, where $\tau = u^{\prime2}/\epsilon$ is the eddy-turnover time.  Both cases reveal a joint PDF this consistent with previous studies of homogeneous turbulence~\cite{cantwell1993behavior}.  Furthermore, the invariants calculated in each simulation corroborate the PDFs of the turbulence statistics in that the overall statistics are captured independent of grid resolution.  Increased grid resolution increases the appearance of outliers and are be associated with intermittency, which should be expected by resolving smaller velocity fluctuations.  \\
\indent Next, three new simulations are introduced with a grid resolution $N=64$ but with different initial $\mathrm{Re}=8.4 \times 10^3, 4.2 \times 10^4$, and $4.2 \times 10^5$.  This is obtained by changing the kinematic viscosity.  All other conditions and forcing remain consistent with the $N=64$ test case at $\mathrm{Re}=6 \times 10^3$.  Figure \ref{fig:forced_re1}(a) shows the longitudinal velocity gradient PDF of the four $N=64$ simulations.  As the Re number increases, the tails of PDF trend closer to the high $\mathrm{Re}_\lambda$ results obtained from DNS and experimental measurements.   Furthermore, Fig. \ref{fig:forced_re1}(b), showing the transverse velocity gradient PDF, suggests similar, yet non-monotonic~\cite{akselvoll1996large}, convergence as the Re number increases.  The higher Re demonstrate that a larger scale separation compared to lower Re is achieved at a minimum state~\cite{zhou2014estimating} such that the turbulence statistics become asymptotic.  The velocity gradient statistics as a function of Re corroborate the statistics as a function of grid resolution (Fig. \ref{fig:forced_pdf1}) to indicate that asymptotic turbulence and dissipation behavior with ILES and VTE scheme can be realized through combination of both high initial Re and sufficient grid resolution.  This ensure that the turbulence will have a sufficient separation of scales.\\
\indent The velocity gradient invariants of the four different Re cases at $N=64$ are shown in Fig. \ref{fig:forced_qr_re}.  These statistics show that as the Re number increases, the joint PDF of the invariants resembles the canonical `teardrop' shape more definitely.  The Q-R plot at higher Re compared to $\mathrm{Re}=6000$ resemble the $N=128$ case shown in Fig. \ref{fig:forced_qr}(b) more as the Re increases.  This further suggests that once a minimum state is reached, statistics become asymptotic.  \\
\indent The balance of terms in enstrophy transport equation introduced in Eqn. (\ref{eqn:resolved_ete}) is shown in Fig. \ref{fig:forced_ens_bal_64}(a) for the $N=64$ case in order to assess the modified equation analysis with forced turbulence.  Similar to the Taylor-Green vortex, the amplification via vortex stretching is balanced by SGS dissipation and diffusion.  Figure \ref{fig:forced_ens_bal_64}(b) shows that the modified equation SGS dissipation and diffusion are consistent with the calculated profile throughout the simulation.  This is a further indication that the implicit model obtained through the modified equation can be used to describe the dissipation of the scheme.  \\
\indent The accuracy of the modified equation SGS dissipation and diffusion at $N=128$ is corroborated in Fig. \ref{fig:forced_ens_bal_128}.   In this case, the enstrophy amplification by vortex stretching is offset by both the SGS dissipation and viscous dissipation.  At this resolution, there is significantly more resolved viscous dissipation than the $N=64$ case.  The calculated SGS dissipation and diffusion is shown to be very similar to the modified equation SGS dissipation and diffusion throughout the simulation. \\
\indent Finally, we examine the average energy spectra over the final half of the simulations 
in Fig. \ref{fig:forced_spec}.  The energy spectra indicate that there is a well-captured low wavenumber regime of all the simulations regardless of the grid resolution.  A zoomed-in plot in the low wavenumber region before the inertial range show that both the $N=256$ and $128$ spectra are nearly converged together, while the modes for the $N=64$ grid resolution approaches similar energy contents and the lowest grid resolutions have slightly higher energy contributions to the modes in this region.  This suggests that very low grid resolution causes slightly more energy to be present in low wavenumber modes. The inertial range may not be wide enough to separate the energy-containing scales from the dissipative scales.  However, the higher resolutions including $N=64$, the resolution becomes asymptotic.
These cases exhibit the formation of an inertial range that is consistent with a $-5/3$ slope.   The energy content at the highest wavenumbers in the $N=256$ cases suggest that the grid resolution is approaching the resolution required for a DNS.  \\
\indent Additional inviscid simulations with the same forcing and initial conditions are performed to quantify the impact of viscosity and resolution.  Figure \ref{fig:forced_spec_re} shows the original forced cases and $\mathrm{Re} = \infty$ cases at grid resolutions of $N=128$ and $64$.   A forced turbulence DNS cases from Ref. ~\cite{rosales2005linear} and explicit LES cases with $N=64$ and $N=128$ from Ref.~\cite{piomelli2015grid} with a similar linear forcing in physical space are included for comparison.  Note that the forcing scheme used in the DNS and explicit SGS cases are performed with the velocity-pressure formulation using the velocity variable while the forcing in physical space employed in the present study uses the vorticity-velocity formulation on the vorticity variable. This may cause parameters selected in the forcing to behave with slight differences between the formulations. However, the spectra reveal that the low wavenumbers are relatively unaffected by the dissipation regardless of the Reynolds number and resolution.  The impact of viscosity is seen in the high wavenumber scales.  Furthermore, the forced ILES simulations show similar behaviors and trends to the DNS and explicit SGS in the low wavenumber region.  The finite Reynolds number does have a larger impact on the inertial range compared to the DNS.  At high Re, there is evidence of asymptotic convergence of flow statistics. 
\subsection{Temporally Evolving Jet}\label{sec:jet}
\begin{figure}
   \begin{center}
      \includegraphics[width=\textwidth]{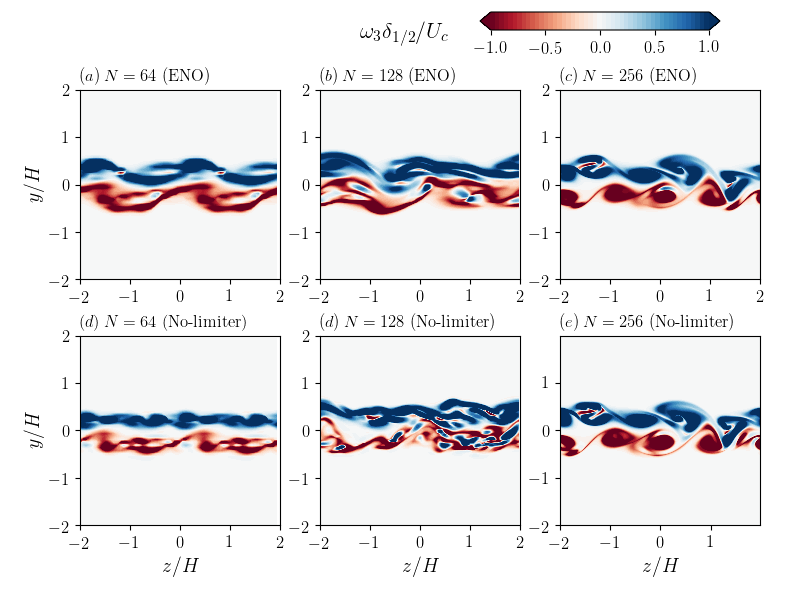}
      \caption{ \label{fig:jet_c} Velocity $\omega_3$ normalized by the jet half-width $\delta_{1/2}$ and centerline velocity for the temporally evolving jet contours for the temporally evolving jet at $t/T_{ref} = 15$. }
   \end{center}
\end{figure}

\begin{figure}
   \begin{center}
      \includegraphics[width=\textwidth]{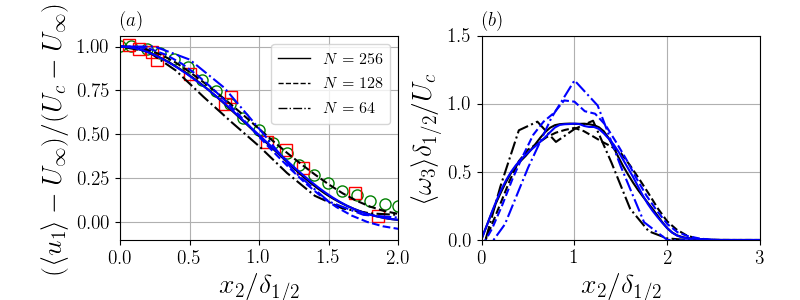}
      \caption{ \label{fig:jet_mean} Profiles of the (a) mean streamwise velocity $\langle u_1 \rangle$ normalized by the centerline velocity $U_c$ compared with experimental results from Ref. \cite{gutmark1976planar} (circles) and Ref. \cite{ramaprian1985lda} (squares) and (b) mean vorticity $\langle\omega_3\rangle$ normalized by the jet half-width $\delta_{1/2}$ and centerline velocity for the temporally evolving jet. Blue - no-limiter, Black - ENO limiter. }
   \end{center}
\end{figure}

\begin{figure}
   \begin{center}
      \includegraphics[width=\textwidth]{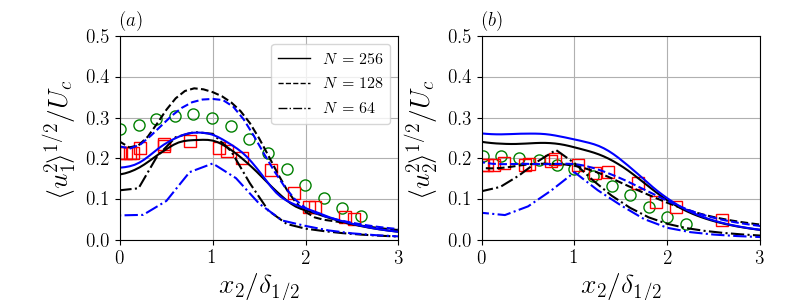}
      \caption{ \label{fig:jet_std} Profiles of the (a) rms (root-mean-square) streamwise velocity $\langle u^2_1 \rangle^{1/2}$ and (b) rms normal velocity $\langle u^2_2 \rangle^{1/2}$ normalized by the centerline velocity $U_c$ compared with experimental results from Ref. \cite{gutmark1976planar} (circles) and Ref. \cite{ramaprian1985lda} (squares). Blue - no-limiter, Black - ENO limiter. }
   \end{center}
\end{figure}

\begin{figure}
   \begin{center}
      \includegraphics[width=\textwidth]{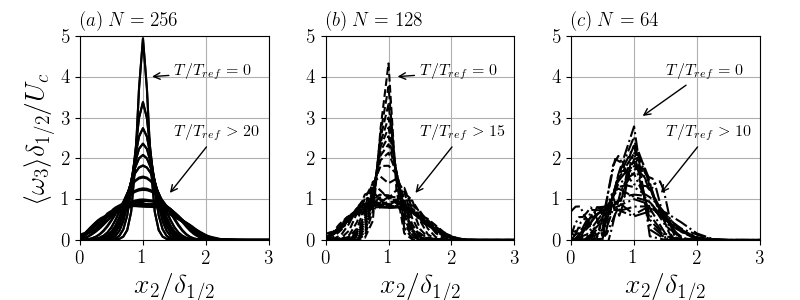}
      \caption{ \label{fig:jet_t} Temporal evolution of the vorticity profiles $\langle\omega_3\rangle$ for (a) $N=256$, (b) $N=128$ and (c) $N=64$.  Each case employs the ENO limiter.}
   \end{center}
\end{figure}

\begin{figure}
   \begin{center}
      \includegraphics[width=\textwidth]{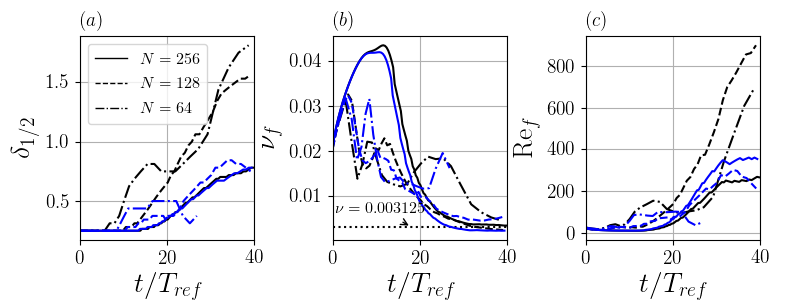}
      \caption{ \label{fig:jet_re} The temporal evolution of the (a) jet half-width, (b) effective viscosity, and (c) effective Re. Blue - no-limiter, Black - ENO limiter. }
   \end{center}
\end{figure}

\begin{figure}
   \begin{center}
      \includegraphics[width=\textwidth]{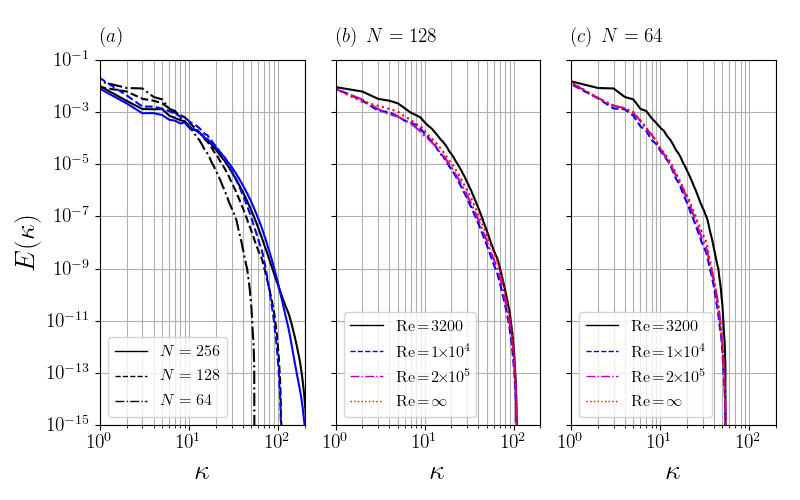}
      \caption{ \label{fig:jet_re_spec} Energy spectra $E(k)$ of simulations temporal simulations: (a)  $\mathrm{Re}=3200$ at several discretizations, Blue - no-limiter, Black - ENO limiter, (b) $N=128$ case at different $\mathrm{Re}$, and (c) $N=64$ case at different $\mathrm{Re}$.}
   \end{center}
\end{figure}

\indent A temporally evolving jet provides a more rigorous test case where both large-scale features (Kelvin-Helmholtz instability) and small-scale fully-developed turbulence are presence.  This is a flow that more closely resembles the complex flows with dominant coherent structures that the present scheme is designed to investigate. \\
\indent The initial velocity flow field from Ref. \cite{da2004effect} is given as follows:
\begin{equation}
    u_1(\bm{x}) = \frac{1}{2} - \frac{1}{2} \tanh \left [ \frac{H}{4\theta_0} \left ( 1 - \frac{2 |x_2|}{H} \right ) \right ]
    \label{eqn:jet_v}
\end{equation}
where $H$ is the initial jet thickness and $\theta_0$ is the initial momentum thickness.  A $H/\theta_0=35$ is selected with a $\textrm{Re} = (U_1 - U_2)H/\nu = 3200$.  The initial $U_1 = 1$ and $U_2=0$. Initial velocity fluctuations are prescribed from the spectrum $E(k) \sim k^4 \exp \left [ -2 (k/k_0)^2 \right ]$.  However, for the present scheme in the vorticity-velocity formulation we impose the initial condition based on the transverse vorticity as follows:
\begin{equation}
   \omega_3(\bm{x}) = \frac{1}{4\theta_0} \frac{|x_2|}{x_2} \cosh^{-2} \left [ \frac{H}{4\theta_0} \left ( 1 - \frac{2 |x_2|}{H} \right ) \right ].
   \label{eqn:jet_w}
\end{equation} 
The initial condition has been shown to be a good estimate of the inlet velocity profile of spatially evolving jets.  However, the temporally evolving jet provides a computationally expedient solution compared to simulating the spatial jet.  
The temporally evolving jet employs periodic boundary conditions on all sides of a cubic domain of $(L_1 \times L_2 \times L_3) = (4H \times 4H \times 4H)$.  Several grid resolution are used for comparison of the effects of the discretization: $N=64, 128, $ and $256$, where $N$ is the number of grid cells per dimension.  At $N=256$, the grid resolution is sufficiently fine to approach a DNS solution similar to the DNS simulation in Ref. \cite{da2008invariants}.  Both a dissipative ENO slope limiter and no-limiter are employed.  The ENO slope limiter is similar to the minmod slope limiter but is not total variation diminishing. Contours of the out-of-plane vorticity are shown for each simulation in Fig. \ref{fig:jet_c} at the time when the flow field becomes starts to become self-similar, $t/T_{ref}=15$, where $T_{ref} = H/\Delta U$, (see Fig. \ref{fig:jet_mean}(a)).  This time is equivalent to $x/H = 7.5$ at the same Re for a spatially evolving jet.  The three different grid resolutions demonstrate how different levels of resolution can affect features in the flow field.  Each case using the ENO limiter show qualitatively the capture of large-scale coherent vortices, while the large-scale structure is less evident with the no-limiter case with low grid resolutions ($N=64$ and $128$). \\
\indent Figure \ref{fig:jet_mean}(a) shows profiles of the mean streamwise velocity $\langle u_1 \rangle$ in $x_2$ direction averaged over the $x_1$ and $x_3$ directions for each grid resolution and both the ENO limiter and no-limiter.  The profiles are normalized by the centerline velocity $U_c = \langle u_1 \rangle(x_2=0)$, while the jet half-width $\delta_{1/2}$, which is calculated with the mean streamwise velocity $\langle u_1 \rangle (x_2=\delta_{1/2}) - U_\infty = \frac{1}{2} \left ( U_c - U_\infty \right ) $ and $U_\infty$ is the streamwise velocity far from the jet, is used to normalize the abscissa.  The profiles are chosen in the regime where the jet wake becomes self-similar.  The grid resolution has a large effect on the amount of time the temporally evolving jet requires to become self-similar and will be discussed below.  The streamwise velocity profiles are compared with several experimental measurements of spatial evolving jet in the self-similar regime.  The present simulation results compare well with experimental measurements and show consistency between the different grid resolutions.  Moreover, the spread in the results are comparable to DNS of temporal jets performed in Refs. \cite{da2004effect, da2008invariants}.  This suggests that even the low grid resolution of $N=64$ with the ENO limiter can capture some of the large-scale features well. The mean transverse vorticity profiles are shown in Fig. \ref{fig:jet_mean}(b) for each simulation in the self-similar regime.  The profiles indicate the mean vorticity from each simulation is captured reasonably well except the $N=64$ with the no-limiter case.  However, the  $N=64$ with dissipative ENO limiter, where the shear layer is only discretized by less than four grid cells, is captured reasonably.  On the other hand, the $N=64$ and $128$ with no-limiter show slightly higher mean vorticity in the shear layer suggesting that under-resolved simulations need additional dissipation.  \\  
\indent The second-order velocity statistics are compared with experimental measurements in Fig. \ref{fig:jet_std}. The rms (root-mean-square) streamwise velocity $\langle u^2_1 \rangle^{1/2}$ profiles, shown in Fig. \ref{fig:jet_std}(a), and rms normal velocity $\langle u^2_2 \rangle^{1/2}$ profiles, shown in Fig. \ref{fig:jet_std}(b), are compared with experimental measurements in the self-similar regime.  While there are some differences between the different test cases, the overall comparison is reasonably well, especially for the lowest grid resolution case with the dissipative ENO limiter, which approximates sufficient dissipation.  On the other hand, the lowest resolution with the no-limiter significantly under predicts the rms velocities.  While scatter in the data can also be attributed to the relatively low number of samples for the temporally evolving jet especially at the lowest grid resolution, the role of numerical dissipation to capture flow statistics is apparent in the jet.   \\
\indent The grid resolution and numerical dissipation affect the temporal evolution of the jet from its initial condition through the transition to turbulence.  Up to this point, the result from the simulations have focused on the flow field after the jet transitions to turbulence, the transient behavior diminishes, and a self-similar regime occurs.  In temporal jet simulations \cite{da2008invariants}, the self-similar regime is obtained at times $t/T_{ref} > 20$ for a similar initialization. This is equivalent to $x/H = 10$ for a spatial-evolving jet.  Figure \ref{fig:jet_t} shows the evolution of the mean transverse vorticity profiles with time for each grid resolution case with the dissipative ENO limiter.  The initial mean vorticity profile across the shear layer is relatively high and sharp. At further instances in time, the vorticity magnitude diminishes to a self-similar solution.  Each grid resolution test case due to the increase numerical dissipation and grid cell size, transitions to the self-similar solution at different times.  For the most resolve case, the self-similar solution is obtained at $t/T_{ref} > 20$, while the lowest grid resolution transitions quicker, around $t/T_{ref} > 10$.  We are effectively solving the flow field at different effective Reynolds numbers through the transition to turbulence but are able to obtain a self-similar solution after the transition. The temporal impact of the numerical scheme on the flow field and the effective Reynolds number is significant when solving temporally evolving flows. \\    
\indent An estimate of the effective Reynolds number is determined from measurable features of the flow field.  It is important to determine the effective Re to determine what the simulation is solving and if it has reached a minimum state in the sense of the asymptotic turbulence statistics.  Here, we assume a high Reynolds number is achieved and using the asymptotic relationship for high Re regime of isotropic turbulence \cite{kaneda2003energy} where $D = \epsilon L / U^3 \approx \frac{1}{2}$ similar to ILES estimates for complex Richtmyer-Meshkov instabilities \cite{zhou2014estimating}.  The velocity scale $U = u^\prime = \langle (\frac{1}{3} u_i u_i)^{1/2} \rangle$ and length scale $L$ is chosen to the jet half-width, which is shown as a function of the time in Fig. \ref{fig:jet_re}(a).  The temporal evolution of the half-width shows that the grid resolution affects the jet spreading, which is expected based Fig. \ref{fig:jet_t}.  
The dissipation is obtained $\epsilon = D U^3/ \delta_{1/2}$ and the effective viscosity is obtain through $\nu_f = \epsilon / \omega_i \omega_i$.  Figure \ref{fig:jet_re}(b) shows the temporal evolution of the effective viscosity.  For each grid resolution and slope limiter case, the initial effective viscosity behaves slightly differently, however, as the jet flow field asymptotically approaches the self-similar solution, the effective viscosity for each simulation approaches a single viscosity, which is the viscosity of the simulation.  The numerical dissipation in the lowest resolution case is able to reach a state such that the asymptotic relationships in the jet can be reached.   
The $\textrm{Re}_f = u^\prime \delta_{1/2}/\nu_f $ is shown in Fig. \ref{fig:jet_re}(c).  It indicates that the lower grid resolution cases with the increased dissipation are in fact solving a slightly different temporally evolving jet problem. However, because the flow comes self-similar at long times, this is not readily noticeable from profiles.  Moreover, experimentation, both numerical and measurement, have shown that at high enough Re based on the inlet velocity and slot width (6000 and Ref. \cite{klein2003investigation} and 1000 in Ref. \cite{deo2008influence}, respectively) the spatially evolving planar jet becomes independent of Re, which comparable to the same Re for the present simulations ($Re \approx 3200$).   \\
\indent Energy spectra of the jet cases are shown in Fig. \ref{fig:jet_re_spec}(a) for $N=256, 128$ and $64$ for both the limited and no-limiter cases at $t/T_{ref} = 40$.  The spectra for the nearly resolved $N=256$ cases are comparatively similar.  Furthermore, the $N=128$ case with the ENO limiter matches reasonably well in the large-scales but is slightly more dissipative than the $N=256$ cases in the high wavenumber range as expected.  The lack of a limiter in the $N=128$ case shows that a non-dissipative scheme can have a large effect on the low wavenumbers.  The lowest resolved cases, $N=64$ show some differences in the low wavenumbers.  Additional Reynolds number cases are performed to assess the low wave number behavior and corroborate the results above that indicate the lower grid resolution are solving a slightly different temporally evolving jet problem.  Figure \ref{fig:jet_re_spec}(b) shows the $N=128$ resolution for $\mathrm{Re} = 3200, 1 \times 10^4, 2 \times 10^5,$ and $\infty$.  The higher Re numbers cases have an asymptotic behavior in the low wavenumbers which converge to the nearly resolved $N=256$ resolution.  Additionally, this behavior is also found in the $N=64$ cases with high Re shown in Fig.  \ref{fig:jet_re_spec}(c). The viscosity $\mathrm{Re}=3200$ plays a significant role in dissipation but can be mitigate by carefully selecting the grid resolution.  The spectra show that with adequate resolution and a high enough Re, the solution converges to correct results especially in the low wavenumber range, which is most important for performing LES.  The estimation of an effective Re becomes especially critical when attempting ILES of more complex flow fields where the flow characteristics can become independent of Re such as wind turbine wakes \cite{chamorro2012reynolds}.  By selecting the grid resolution for the problem, an ILES with the present scheme can be employed such that the large-scale structures are captured and small-scale turbulence are implicitly modeled to obtain physically accurate, expedient results.

%
\section{Conclusions}\label{sec:conclusions}
\indent The subgrid scale characteristics and effectiveness of an upwind finite volume scheme for the vorticity transport equations were investigated.  The numerical scheme employs a generalized Riemann problem-based multi-dimensional wave propagation approach.  Modified equation analysis was used to characterize the dissipation and backscatter. The analysis revealed two limits for including dissipation implicitly through numerics: 1.) a low dissipation limit using a second-order central difference, most appropriate in smooth areas, i.e. regions dominated by large vortical coherent structures; and 2.) a high dissipation limit using a first order upwind difference, used when the vorticity changes rapidly across grid cells, i.e., regions of under-resolved turbulence.  While the former is ideal in well-resolved areas of the flow field, the latter is necessary in regions in which dissipation is essential to account for the transfer of energy from the resolved scales to the sub-grid scales in the absence of an explicit SGS model.  
The modified equation at the high dissipation limit contains many terms, some of which can be combined into forms that are similar to commonly used explicit sub-grid scale models including the tensor-gradient models, hyper-viscosity model, and a simple gradient model.  This serves a qualitative tool to understand the implications of using the present scheme for ILES.  We are careful to note that---for ILES in general---the terms in the modified equation do not have to be similar to explicit SGS terms.  Rather, the similarities to known models  provide insight in characterizing the scheme. \\
\indent To characterize turbulence with the present scheme for the ILES methodology, a series of turbulence-in-a-box simulations of the Taylor-Green vortex was performed to understand the process of energy transfer from energy-containing scales to the sub-grid scale.   The Taylor-Green vortex cases revealed that the grid resolution must be carefully taken into account in order to obtain desired results.    
The grid resolution is the largest factor for the magnitude of the effective Re while the choice of limiter, which subtly controls the numerical dissipation, has an impact on the accuracy.   However, in under-resolved simulations, a dissipative limiter is essential.  
The dissipation terms obtained from the modified equation analysis are shown to faithfully represent the implicit SGS torque in the Taylor-Green vortex case. In high Reynolds number flows, where there is a marked separation of scales, the method is able to represent the high energy modes.   Further numerical experiments with forced turbulence revealed that high-Reynolds number asymptotic turbulence statistics can be reasonably captured with the ILES methodology for this vorticity-velocity formulation scheme. \\
\indent Finally, simulations of a temporally evolving jet, which contains both large-scale vortical structures and fine-scale turbulence show that under-resolved numerics can capture asymptotic turbulence statistics and large-scale features.  The method is particularly useful when the effective Reynolds number is past a threshold beyond which the flow is dependent on the Reynolds number.  

The simulations studied herein represent canonical flows which allow us to build our understanding of the method to more complex flows.  The simulation tests show that  coarse grid resolutions provide a good estimate for large energy-containing modes given a large enough inertial range.  This particular vorticity-velocity scheme was designed to capture and preserve large vorticity structures in flows in which fully-developed small-scale turbulence tends to localized and large energy-containing structures dominate the flow. Our previous work \cite{foti2018multi} showed how the scheme can be used to capture vortical structures, while this work indicates that the impact of fine-scale turbulence on the energy-containing scales can also be reasonably represented.   
%
%
\section*{Acknowledgments}
This work is supported through a subcontract from Continuum Dynamics, Inc. under Navy STTR Phase II contract N68335-17-C-0158 entitled ``Advanced Wake Turbulence Modeling for Naval CFD Applications'', with guidance from Dr. Glen Whitehouse at Continuum Dynamics, Inc., and Drs. David Findlay and James Forsythe at the Naval Air Systems Command. 
 
This paper was prepared as an account of work sponsored by an agency of the United States (U.S.) Government.  Neither the U.S. Government nor any agency thereof, nor any of their employees, makes any warranty, express or implied, or assumes any legal liability or responsibility for the accuracy, completeness, or usefulness of any information, apparatus, product, or process disclosed, or represents that its use would not infringe privately owned rights.  Reference herein to any specific commercial product, process, or service by trademark, manufacturer, or otherwise does not necessarily constitute or imply its endorsement, recommendation, or favoring by the U.S. Government or any agency thereof.  The views and opinion expressed herein do not necessarily state or reflect those of the U.S. Government or any agency thereof.
%
%
%
\bibliography{bibl}
\end{document}